\documentclass[12pt]{article}
\pdfoutput=1
\usepackage{amssymb}
\usepackage{epsfig}

\usepackage{graphicx}

\def\L{\mathcal L}
\def\e{\varepsilon}

\newcommand{\wt}{\widetilde}

\begin{document}

\def\a{\alpha}
\def\b{\beta}
\def\c{\chi}
\def\d{\delta}
\def\e{\epsilon}
\def\f{\phi}
\def\g{\gamma}
\def\h{\eta}
\def\i{\iota}
\def\j{\psi}
\def\k{\kappa}
\def\l{\lambda}
\def\m{\mu}
\def\n{\nu}
\def\o{\omega}
\def\p{\pi}
\def\q{\theta}
\def\r{\rho}
\def\s{\sigma}
\def\t{\tau}
\def\u{\upsilon}
\def\x{\xi}
\def\z{\zeta}
\def\D{\Delta}
\def\F{\Phi}
\def\G{\Gamma}
\def\J{\Psi}
\def\L{\Lambda}
\def\O{\Omega}
\def\P{\Pi}
\def\Q{\Theta}
\def\S{\Sigma}
\def\U{\Upsilon}
\def\X{\Xi}

\def\ve{\varepsilon}
\def\vf{\varphi}
\def\vr{\varrho}
\def\vs{\varsigma}
\def\vq{\vartheta}

\def\dg{\dagger}                                     
\def\ddg{\ddagger}                                   
\def\wt#1{\widetilde{#1}}                    
\def\mt{\widetilde{m}_1}
\def\mti{\widetilde{m}_i}
\def\rt{\widetilde{r}_1}
\def\mtt{\widetilde{m}_2}
\def\mttt{\widetilde{m}_3}
\def\rtt{\widetilde{r}_2}
\def\mb{\overline{m}}
\def\VEV#1{\left\langle #1\right\rangle}        
\def\be{\begin{equation}}
\def\ee{\end{equation}}
\def\ds{\displaystyle}
\def\ra{\rightarrow}

\def\bea{\begin{eqnarray}}
\def\eea{\end{eqnarray}}
\def\NO{\nonumber}
\def\Bar#1{\overline{#1}}


\def\pl#1#2#3{Phys.~Lett.~{\bf B {#1}} ({#2}) #3}
\def\np#1#2#3{Nucl.~Phys.~{\bf B {#1}} ({#2}) #3}
\def\prl#1#2#3{Phys.~Rev.~Lett.~{\bf #1} ({#2}) #3}
\def\pr#1#2#3{Phys.~Rev.~{\bf D {#1}} ({#2}) #3}
\def\zp#1#2#3{Z.~Phys.~{\bf C {#1}} ({#2}) #3}
\def\cqg#1#2#3{Class.~and Quantum Grav.~{\bf {#1}} ({#2}) #3}
\def\cmp#1#2#3{Commun.~Math.~Phys.~{\bf {#1}} ({#2}) #3}
\def\jmp#1#2#3{J.~Math.~Phys.~{\bf {#1}} ({#2}) #3}
\def\ap#1#2#3{Ann.~of Phys.~{\bf {#1}} ({#2}) #3}
\def\prep#1#2#3{Phys.~Rep.~{\bf {#1}C} ({#2}) #3}
\def\ptp#1#2#3{Progr.~Theor.~Phys.~{\bf {#1}} ({#2}) #3}
\def\ijmp#1#2#3{Int.~J.~Mod.~Phys.~{\bf A {#1}} ({#2}) #3}
\def\mpl#1#2#3{Mod.~Phys.~Lett.~{\bf A {#1}} ({#2}) #3}
\def\nc#1#2#3{Nuovo Cim.~{\bf {#1}} ({#2}) #3}
\def\ibid#1#2#3{{\it ibid.}~{\bf {#1}} ({#2}) #3}

\title{
\vspace*{15mm}
\bf Strong thermal leptogenesis \\ and the \\ absolute neutrino mass scale}
\author{{\Large Pasquale Di Bari, Sophie E. King, Michele Re Fiorentin}
\\
{\it  School of Physics and Astronomy}, \\
{\it University of Southampton,} 
{\it  Southampton, SO17 1BJ, U.K.}
}

\maketitle \thispagestyle{empty}

\vspace{-10mm}

\begin{abstract}
We show that successful strong thermal leptogenesis, where  
the final asymmetry is independent of the initial conditions and in particular a 
large pre-existing asymmetry is efficiently washed-out, 
favours values of the  lightest neutrino mass $m_1  \gtrsim  10\,{\rm meV}$ for 
normal ordering (NO) and $m_1 \gtrsim 3\,{\rm meV}$ for inverted ordering (IO)
for models with orthogonal matrix entries respecting $|\O_{ij}^2| \lesssim 2$.
We show analytically why lower values of $m_1$ require a higher level of fine tuning in the seesaw formula
and/or in the flavoured decay parameters (in the electronic for NO,  in the muonic for IO).
 We also show how this constraint exists thanks to the measured values of the neutrino mixing angles
and could be tightened by a future determination of the Dirac phase. 
Our analysis also allows us to place a more stringent constraint for a specific model or class of models, such as $SO(10)$-inspired 
models, and shows that some models cannot realise strong thermal leptogenesis for any value of $m_1$. 
A scatter plot analysis fully supports the analytical results. 
We also briefly discuss the interplay with absolute neutrino mass scale experiments 
concluding that they will be able in the coming years to either corner  strong thermal leptogenesis 
or find positive signals pointing to a non-vanishing $m_1$.
Since the constraint is much stronger for NO than for IO, it is very important that
new data from planned neutrino oscillation experiments will be able to  solve the ambiguity.
\end{abstract}
\newpage
\section{Introduction}

The observed matter-antimatter asymmetry of the  Universe is a long standing cosmological puzzle
calling for physics beyond the Standard Model (SM). 
In terms of the baryon-to-photon number ratio $\eta_B$ the matter-antimatter asymmetry is today accurately and precisely measured by CMB observations. Recently the {\em Planck} collaboration   found from CMB anisotropies plus lensing data 
\footnote{More precisely the Planck collaboration finds
$\O_{B}\,h^2=0.02217 \pm 0.00033$ corresponding to $10^{10}\,\eta_B \simeq 273.6\,\O_B\,h^2 \simeq
6.065\pm 0.09$.}
\cite{planck}
\be
\eta_B^{\rm CMB} = (6.1 \pm 0.1) \times 10^{-10} \,  .
\ee
Leptogenesis \cite{fy} provides an attractive solution since it relies on 
a minimal and natural way to extend the SM  incorporating   
neutrino masses and mixing  discovered in neutrino oscillation experiments:
the seesaw mechanism \cite{seesaw}.  
At the same time it should be noticed that leptogenesis
also relies on the Brout-Englert-Higgs mechanism and, therefore, the recent discovery of the 
Higgs boson at the LHC nicely contributes to support the picture. 
On the other hand the non-observation of new physics at the LHC so far, 
places stronger constraints on low scale baryogenesis scenarios such as, for example, electroweak
baryogenesis within the minimal supersymmetric standard model \cite{nardini}.  
 

The prediction of the baryon asymmetry relies on some assumption on the initial conditions.  
A plausible and common one is that an inflationary stage before leptogenesis 
resets the initial conditions in the early Universe, enforcing vanishing values of the asymmetry and of the 
right-handed (RH) neutrino abundances prior to  the onset of leptogenesis. 
However, it cannot be excluded, especially at the high temperatures required by
a minimal scenario of leptogenesis \cite{review}, that other mechanisms, 
such as gravitational \cite{grav},   GUT  \cite{GUT}, Affleck-Dine baryogenesis \cite{AD},
generate a large asymmetry at the end of inflation and/or prior to the onset of leptogenesis. 

Since these mechanisms escape experimental probes,
it would be certainly more attractive  if the final asymmetry from 
leptogenesis were independent of the initial conditions.  
In this paper we show that, given the current low energy neutrino data, the possibility to enforce 
independence of the initial conditions in leptogenesis, so called strong thermal leptogenesis, 
 barring quasi-degenerate RH neutrino masses
and strong fine tuned cancellations in the  flavoured decay parameters and in the seesaw formula, 
implies a lower bound on the absolute neutrino mass scale, more specifically on the lightest neutrino mass.  
Though this lower bound can be evaded allowing for fine tuned cancellations, most 
of the models require values of the lightest neutrino mass that will be tested during the coming years,
especially in the case of NO.

The plan of the paper is the following. In Section 2 we introduce some basic notation and review
current experimental information on low energy neutrino parameters. In Section 3
we briefly discuss strong thermal leptogenesis. In Section
4 we show the existence of a lower bound on the neutrino masses under certain conditions.
We also present results from a scatter plot analysis confirming the existence of the lower bound and
at the same time showing how the bulk of models require values of the lightest neutrino mass 
that can be potentially tested in future years mainly with cosmological observations.  
In Section 5 we draw the conclusions. 

\section{General set up}

We assume a minimal model of leptogenesis  where the SM Lagrangian is extended
introducing three RH neutrinos $N_i$ with Yukawa couplings $h$ and a Majorana mass term $M$.
After spontaneous symmetry breaking the Higgs vev generates a Dirac neutrino mass term $m_D$.
In the seesaw limit the spectrum of neutrino masses splits into a set of three heavy neutrinos
with masses $M_1 \leq M_2 \leq M_3$, approximately equal to the eigenvalues of $M$, and into a
set of light neutrinos with masses $m_1 \leq m_2 \leq m_3$ given by the seesaw formula
\be
D_m =  U^{\dagger} \, m_D \, {1\over D_M} \, m_D^T  \, U^{\star}  \,   ,
\ee
written in a basis where both the Majorana mass and the charged lepton mass matrices are diagonal,
so that $U$ can be identified with the PMNS leptonic mixing matrix. 

From neutrino oscillation experiments we know two mass squared differences, $\Delta m^2_{\rm atm}$
and $\Delta m^2_{\rm sol}$. Neutrino masses can then be either NO,  with
$m_3^{\, 2} - m_2^{\, 2} = \Delta m^2_{\rm atm}$ and $m^2_2 - m^2 _1 = \Delta m^2_{\rm sol}$, 
or IO, with $m_3^2 - m_2^2 = \Delta m^2_{\rm sol}$ and $m^2_2 - m^2 _1 = \Delta m^2_{\rm atm}$.
For example, in a recent global analysis \cite{valle}, and analogously in \cite{newfogli,gonzalez}, it is found
$m_{\rm atm}\equiv \sqrt{m^{\,2}_3 - m_1^{\,2}} \simeq 0.0505\,(0.0493)\,{\rm eV}$
and  $m_{\rm sol}\equiv \sqrt{\Delta m^2_{\rm sol}} \simeq 0.0087\,{\rm eV}$.

In order to fix completely the three light neutrino masses,
there is just one parameter left   to be measured, the
so called absolute neutrino mass scale.
This can be conveniently parameterised in terms of  the lightest neutrino mass $m_1$. 
The most stringent upper bound on $m_1$ comes from cosmological observations. 
A conservative upper bound on the sum of the neutrino masses has been recently placed by the {\em Planck} collaboration \cite{planck}. Combining {\em Planck} and high-${\ell}$ CMB anisotropies, WMAP polarisation  and baryon acoustic oscillation data, it is found   $\sum_i \, m_i \lesssim 0.23\,{\rm eV}\,(95\% {\rm C.L.})$.  
When neutrino oscillation results are combined, 
this translates into an upper bound on the lightest neutrino mass,
\be\label{PLANCK}
m_1 \lesssim 0.07 \, {\rm eV}  \;\; (95\% \, \mbox{\rm C.L.}) \,  ,
\ee  
showing how cosmological observations  
start to corner quasi-degenerate neutrinos. 

For NO the leptonic mixing matrix can be parameterised as 
\begin{equation}\label{Umatrix}
U^{\rm (NO)}=\left( \begin{array}{ccc}
c_{12}\,c_{13} & s_{12}\,c_{13} & s_{13}\,e^{-{\rm i}\,\d} \\
-s_{12}\,c_{23}-c_{12}\,s_{23}\,s_{13}\,e^{{\rm i}\,\d} &
c_{12}\,c_{23}-s_{12}\,s_{23}\,s_{13}\,e^{{\rm i}\,\d} & s_{23}\,c_{13} \\
s_{12}\,s_{23}-c_{12}\,c_{23}\,s_{13}\,e^{{\rm i}\,\d}
& -c_{12}\,s_{23}-s_{12}\,c_{23}\,s_{13}\,e^{{\rm i}\,\d}  &
c_{23}\,c_{13}
\end{array}\right)
\, {\rm diag}\left(e^{i\,\rho}, 1, e^{i\,\sigma}
\right)\, ,
\end{equation}
($s_{ij}\equiv \sin\theta_{ij}, c_{ij} \equiv \cos \theta_{ij}$) while for IO, within our convention of labelling  light neutrino masses,
the columns of the leptonic mixing matrix have to be permuted in a way that
\be
U^{\rm (IO)} =  \left( \begin{array}{ccc}
s_{13}\,e^{-{\rm i}\,\d} & c_{12}\,c_{13} & s_{12}\,c_{13}  \\
s_{23}\,c_{13} & -s_{12}\,c_{23}-c_{12}\,s_{23}\,s_{13}\,e^{{\rm i}\,\d} &
c_{12}\,c_{23}-s_{12}\,s_{23}\,s_{13}\,e^{{\rm i}\,\d} \\
c_{23}\,c_{13} & s_{12}\,s_{23}-c_{12}\,c_{23}\,s_{13}\,e^{{\rm i}\,\d}
& -c_{12}\,s_{23}-s_{12}\,c_{23}\,s_{13}\,e^{{\rm i}\,\d}
\end{array}\right)
\, {\rm diag}\left(e^{i\,\sigma}, e^{i\,\rho}, 1  \right) \,  .
\ee
The mixing angles, respectively the reactor, the solar and the atmospheric one, 
are now constrained within the following $1\s$ ($3\s$) ranges \cite{newfogli} for NO and IO respectively,
\bea\label{mixinganglesNO}
s^2_{13} & = & 0.0234^{+0.0022}_{-0.0018} \,  
                \;\; (0.0177\mbox{--}0.0297) \,  
                 \;\; \mbox{\rm and} \;\;
                s^2_{13} = 0.0239^{+0.0021}_{-0.0021} \,  
                \;\; (0.0178\mbox{--}0.0300) \,
                 \,  ,\\ \nonumber
s^2_{12}& = & 0.308\pm 0.017 \,  
                \;\; (0.259\mbox{--}0.359) \,  
                \;\; \mbox{\rm and} \;\;
             s^2_{12} =   0.308\pm 0.017 \,  
                \;\; (0.259\mbox{--}0.359) \,  ,
                 \\ \nonumber
s^2_{23} & = & 0.425^{+0.029}_{-0.027} \,  
                \;\; (0.357\mbox{--}0.641) \,  
                \;\; \mbox{\rm and} \;\;
          s^2_{23} =  0.437^{+0.059}_{-0.029} \oplus 0.531\mbox{--}0.610 \,  
                \;\; (0.363\mbox{--}0.659) \, 
                 \,  .
\eea
It is interesting that current experimental data also start to put constraints on the
Dirac phase and the following best fit values and $1\s$ errors are found  for NO and IO respectively,
\be
\d/\pi = -0.61^{+0.33}_{-0.27} \,\,  
\;\;\mbox{\rm and} \;\;
\d/\pi = -0.65^{+0.24}_{-0.39} \,  ,
\ee
while all values $[-\pi,+\pi]$ are still allowed at $3\,\s$.
\footnote{It is also useful to give the constraints on the angles and on $\d$ in degrees:
\bea
\theta_{13} & = &  8.8^{\circ}\pm 0.4^{\circ} \, \;\;  (7.6^{\circ}\mbox{--}9.9^{\circ}) 
 \;\; \mbox{\rm and} \;\;
 \theta_{13}  =    8.9^{\circ}\pm 0.4^{\circ} \, \;\; (7.7^{\circ}\mbox{--}10^{\circ}) \,  ,
 \\ \nonumber
\theta_{12} & = &  33.70^{\circ}\pm 1.05^{\circ} \,  \;\;  (30.6^{\circ}\mbox{--}36.8^{\circ}) 
\;\; \mbox{\rm and} \;\;
\theta_{12}  =   33.7^{\circ}\pm 1.1^{\circ} \,  \;\;  (30.6^{\circ}\mbox{--}36.8^{\circ}) \,  , 
\\ \nonumber
\theta_{23} & = &  40.7^{\circ}\pm 1.6^{\circ} \,  \;\;  (36.7^{\circ}\mbox{--}53.2^{\circ}) 
\;\; \mbox{\rm and} \;\;
\theta_{23}  =   
                 {41.4^{\circ}}^{+3.4^{\circ}}_{-1.8^{\circ}} \oplus 46.8^{\circ} \mbox{--} 51.3^{\circ}\, 
                  \;\;  (37^{\circ}\mbox{--}54.3^{\circ})  \,  , 
\\ \nonumber
\d  & = &  -{110^{\circ}}^{+59^{\circ}}_{-49^{\circ}}
\;\; \mbox{\rm and} \;\;
\d    =  -{117^{\circ}}^{+43^{\circ}}_{-70^{\circ}} \,  .
 \eea  }

\section{Strong thermal leptogenesis and the $N_2$-dominated scenario}

Within an unflavoured scenario and assuming, conservatively, that only the lightest RH neutrinos
thermalise, the strong thermal condition translates quite straightforwardly into a
condition on the lightest RH neutrino decay parameter $K_1 \equiv \widetilde{\Gamma}_1/H(T=M_1)$,
where $H$ is the expansion rate and $\widetilde{\Gamma}_1$ is the $N_1$ total decay width. 
 Given a pre-existing asymmetry $N_{B-L}^{\rm p,i}$, the relic value after the lightest 
RH neutrino wash-out is simply given by \cite{fy,window}
\be
N_{B-L}^{\rm p,f} = e^{-{3\pi \over 8}\, K_1} \,  N_{B-L}^{\rm p,i}  \,  ,
\ee
where we are indicating with $N_{X}$ the abundance of any (extensive) quantity $X$ in a co-moving volume containing
one RH neutrino in ultra-relativistic thermal equilibrium 
(so that $N^{\rm eq}_{N_1}(T\gg M_1) = 1$).
The relic value of the pre-existing asymmetry would then result in a contribution to $\eta_B$ given by
$\eta_B^{\rm p} \simeq 0.01 \, N_{B-L}^{\rm p, f}$,
taking into account the dilution due to photon production and the sphaleron conversion coefficient. 

Imposing $|\eta_B^{\rm p}| \lesssim 0.1\,\eta_B^{\rm lep} \simeq 0.1\, \eta_B^{\rm CMB}$, where $\eta_{B}^{\rm lep}$ is the contribution  coming from leptogenesis,  immediately yields the simple condition 
$K_1 \gtrsim K_{\rm st}(N_{B-L}^{\rm p,i}) $, with
\be\label{Kstrong}
K_{\rm st}(x)\equiv {8\over 3\pi }\,\left[\ln \left({0.1 \over \eta_B^{\rm CMB}}\right) + \ln |x|\right] \simeq 16 + 0.85\,\ln |x| \,   ,
\ee 
where $|N_{B-L}^{\rm p,i}|$  is assumed to be large,  meaning that 
$|N_{B-L}^{\rm p,i}| \gg 100 \, \eta_B^{\rm CMB} \sim 10^{-7}$. Since  $K_1\geq m_1/m_{\star}$, 
where $m_{\star} \simeq 1.1 \times 10^{-3}\,{\rm eV}$, 
the requirement $m_1 \gtrsim 10^{-3}\,\,K_{\rm st}\,{\rm eV}$   is
a sufficient (but not necessary)  condition for strong thermal leptogenesis.

When flavour effects are considered, the possibility to satisfy both successful leptogenesis,  
$\eta_B^{\rm lep} \simeq \eta_B^{\rm CMB}$, and  strong thermal condition, $|\eta_B^{\rm p}| 
\lesssim 0.1\,\eta_B^{\rm lep}$, relies on much more restrictive 
conditions \cite{problem},  due to the 3-dim flavour space and to the fact that the 
RH neutrino wash-out acts only along a specific flavour component \cite{strumianardinir}.

It is then possible to show \cite{problem} that only in a {\em $N_2$-dominated scenario} \cite{geometry}, 
defined by having $M_1 \ll 10^9\,$GeV and $M_2 \gtrsim 10^9\,{\rm GeV}$,
so that the observed asymmetry is dominantly produced by the $N_2$ RH neutrinos,
with the additional requirements $M_2 \lesssim 5 \times 10^{11}\,{\rm GeV}$  
\footnote{In this way the asymmetry production from $N_2$
decays occurs in the two-flavour regime \cite{strumianardinir,densitymatrix}.}  
and that the asymmetry is dominantly produced in the tauon flavour, 
one can have successful strong thermal leptogenesis.  

In the $N_2$-dominated scenario the contribution to the asymmetry from 
leptogenesis can  be calculated as  the sum of the 
three (charged lepton) flavoured asymmetries $\D_\a \equiv B/3 - L_\a$,
\cite{N2dominated,fuller,densitymatrix2}
\bea\label{twofl} \nonumber
N_{B-L}^{\rm lep, f} & \simeq &
\left[{K_{2e}\over K_{2\tau_2^{\bot}}}\,\ve_{2 \tau_2^{\bot}}\kappa(K_{2 \tau_2^{\bot}}) 
+ \left(\ve_{2e} - {K_{2e}\over K_{2\tau_2^{\bot}}}\, \ve_{2 \tau_2^{\bot}} \right)\,\kappa(K_{2 \tau_2^{\bot}}/2)\right]\,
\, e^{-{3\pi\over 8}\,K_{1 e}}+ \\ \nonumber
& + &\left[{K_{2\mu}\over K_{2 \tau_2^{\bot}}}\,
\ve_{2 \tau_2^{\bot}}\,\kappa(K_{2 \tau_2^{\bot}}) +
\left(\ve_{2\mu} - {K_{2\mu}\over K_{2\tau_2^{\bot}}}\, \ve_{2 \tau_2^{\bot}} \right)\,
\kappa(K_{2 \tau_2^{\bot}}/2) \right]
\, e^{-{3\pi\over 8}\,K_{1 \mu}}+ \\
& + &\ve_{2 \tau}\,\kappa(K_{2 \tau})\,e^{-{3\pi\over 8}\,K_{1 \tau}} \,  ,
\eea
where $K_{2\tau_2^{\bot}} \equiv K_{2e} + K_{2\mu}$ and 
$\ve_{2\tau_2^{\bot}} \equiv \ve_{2e} + \ve_{2\mu}$.
As we will show soon, the strong thermal condition implies $K_{1e}, K_{1\m} \gg 1$ and, therefore, 
in this case the contribution to the asymmetry from leptogenesis 
simply reduces to 
\be
N_{B-L}^{\rm lep, f} \simeq \ve_{2 \tau}\,\kappa(K_{2 \tau})\,e^{-{3\pi\over 8}\,K_{1 \tau}}\,  .
\ee
The baryon-to-photon number ratio from leptogenesis can then be simply calculated
as $\eta_B^{\rm lep} \simeq 0.01\,N_{B-L}^{\rm lep,f}$.
The flavoured decay parameters $K_{i\a}$ are defined as
\be
K_{i\a}\equiv {\G_{i\a}+\overline{\G}_{i\a}\over H(T=M_i)}= 
{|m_{D\a i}|^2 \over M_i \, m_{\star}} \,  .
\ee
The $\Gamma_{i\a}$'s and the $\bar{\Gamma}_{i \a}$'s can be regarded
as the zero temperature limit of the flavoured decay rates into $\a$ leptons, 
$\Gamma (N_i \ra \phi^\dagger \, l_\alpha)$,
and anti-leptons, $\Gamma (N_i \ra \phi \, \bar{l}_\alpha)$ in a three-flavoured regime, where
lepton quantum states can be treated as an incoherent mixture  of the three flavour components.  
They are related to the total decay widths by  $\widetilde{\G}_i = \sum_\alpha \, \widetilde{\G}_{i\a}$, 
with $\widetilde{\G}_{i\a} \equiv \G_{i\a} + \bar{\G}_{i \a}$. 
The efficiency factors can be calculated using \cite{pedestrians,flavorlep}
\be\label{kappa}
\k(K_{2\a}) = 
{2\over z_B(K_{2\a})\,K_{2\a}}
\left(1-e^{-{K_{2\a}\,z_B(K_{2\a})\over 2}}\right) \,  , \;\; z_B(K_{2\a}) \simeq 
2+4\,K_{2\a}^{0.13}\,e^{-{2.5\over K_{2\a}}} \,   .
\ee
This is the expression for an initial thermal abundance but, since we 
will impose the strong thermal leptogenesis condition,  this will automatically select
the region of the space of parameters where there is no dependence on 
the initial conditions anyway. 
\footnote{Moreover in this case this analytical expression approximates the numerical result
with an error below $10\%$.}

Within the $N_2$-dominated scenario the flavoured $C\! P$ asymmetries,  defined as 
$\ve_{2\a} \equiv (\overline{\G}_{2\a}-\G_{2\a})/(\G_{2\a}+\overline{\G}_{2\a})$, 
can be calculated in the hierarchical limit simply using \cite{crv}
\be\label{eps2a}
\ve_{2\a} \simeq
\overline{\ve}(M_2) \,  \b_{2\a} \, ,  \;\;\;
\b_{2\a} \equiv {{\rm Im}\left[m_{D\a 2}^{\star}
m_{D\a 3}(m_D^{\dag}\, m_D)_{2 3}\right]\over M_2\,M_3\,\mtt\,m_{\rm atm} }\,   ,
\ee
with $\mtt \equiv (m_D^{\dag}\, m_D)_{2 2}/M_2$ and
$\overline{\ve}(M_2) \equiv {[3 / (16\,\pi)]} \, {(M_2\,m_{\rm atm} / v^2)}$.

In the orthogonal parameterisation the neutrino Dirac mass matrix, in the basis where 
both charged lepton and RH neutrino mass matrices are diagonal, can be written as 
$m_D =  U\,\sqrt{D_m}\, \Omega \, \sqrt{D_M}$\, , where $\O$ is an orthogonal matrix 
encoding the information on the properties of the  RH neutrinos \cite{casas}. 
This parameterisation is quite convenient in order to easily account for 
the experimental low energy neutrino  information.  Barring strong cancellations in the seesaw formula, 
one typically expects $|\O^2_{ij}| \lesssim {\cal O}(1)$. More generally, we will impose a condition
$|\O^2_{ij}| < M_{\Omega}$, studying the dependence of the results on $M_{\O}$.   
In the orthogonal parametrisation  the flavoured decay parameters can be calculated as
\be\label{fdp}
K_{i\a} = \left|\sum_j\,\sqrt{m_j\over m_{\star}}\,U_{\a j}\,\O_{j i}\right|^2 \,  .
\ee
The quantity $\b_{2\a}$ can also  be expressed in the orthogonal 
parameterisation,
\be
\b_{2\a} =   {\rm Im} \Big[ \sum\limits_{k,h,l}
{m_{k}\,\sqrt{m_{h}\,m_{l}} \over \mtt \, m_{\rm atm}}\,
\,\O^*_{k2}\,\O_{k3}\,\O^*_{h2}\,\O_{l 3}\,U_{\alpha h}^* \, U_{\alpha l} \Big] \,   .
\ee 
Now, we have finally to impose the strong thermal condition, and to this extent 
we need to calculate  the relic value of the pre-existing asymmetry distinguishing 
two different cases. 

\subsection{Case $M_3 \gtrsim 5\times 10^{11}\,{\rm GeV}$}

In the case $M_3 \gtrsim 5 \times 10^{11}\,{\rm GeV},$ the heaviest RH neutrino either, for $M_3 \gg T_{RH}$, 
is not thermalised  or it cannot in general wash-out completely the pre-existing asymmetry, as requested by the strong thermal leptogenesis condition. This is because the wash-out would occur in the one-flavour regime and, for a generic pre-existing asymmetry,  the component orthogonal to the $N_3$-flavour direction would survive.  
Therefore,   without any loss of generality, we can simply neglect its presence.  
The relic value of the pre-existing asymmetry can then be calculated as \cite{strongSO10lep}
$N_{B-L}^{\rm p, f} = \sum_{\a} \,  N_{\D_\a}^{\rm p,f}$ \, ,  with
\bea\label{finalpas}
N_{\D_\t}^{\rm p,f} & = & 
(p^0_{{\rm p}\t}+\D p_{{\rm p}\tau})\,  e^{-{3\pi\over 8}\,(K_{1\t}+K_{2\t})} \, N_{B-L}^{\rm p,i} \,  , \\  \nonumber
N_{\D_\m}^{\rm p,f} & = & 
\left\{(1-p^0_{{\rm p}\t})\,\left[
p^0_{\mu\t_2^{\bot}}\, p^0_{{\rm p}\t^\bot_2}\,
e^{-{3\pi\over 8}\,(K_{2e}+K_{2\m})} + (1-p^0_{\m\t_2^{\bot}})\,(1-p^0_{{\rm p}\t^\bot_2}) \right]  + 
\D p_{{\rm p}\mu}\right\}
\,e^{-{3\pi\over 8}\,K_{1\m}}\, N_{B-L}^{\rm p,i}
 ,  \\  \nonumber
N_{\D_e}^{\rm p,f}& = & 
\left\{(1-p^0_{{\rm p}\t})\,\left[ 
p^0_{e\t_2^{\bot}}\,p^0_{{\rm p}\t^\bot_2}\,
e^{-{3\pi\over 8}\,(K_{2e}+K_{2\m})} + (1-p^0_{e \t_2^{\bot}})\,(1-p^0_{{\rm p}\t^\bot_2}) \right]  + \D p_{{\rm p} e}\right\}
 \,e^{-{3\pi\over 8}\,K_{1e}} \,\, N_{B-L}^{\rm p,i} \,   .
\eea
In this expression  
\footnote{Notice that in the limit $K_{1\a} = K_{2\a}= 0\, (\a=e,\m,\t) $
one has $\sum_\a \ N^{\rm p,f}_{\D_\a}=N^{\rm p,i}_{B-L}$. Notice also that this expression 
incorporates flavour projection \cite{strumianardinir} 
and exponential suppression of the parallel components, 
two effects that have been both confirmed within  a 
density matrix approach \cite{densitymatrix2}.}
the quantities $p^0_{{\rm p} \tau}$ and $p^0_{{\rm p}\tau_2^{\bot}}$ 
are the fractions of the pre-existing asymmetry in the tauon and $\tau_2^{\bot}$ 
components respectively, where $\tau_2^{\bot}$ is the $\tau$-orthogonal flavour component of the leptons
produced by $N_2$ decays, while 
$p^0_{\alpha \t_2^{\bot}} \equiv K_{2\alpha}/(K_{2e}+K_{2\mu}) \; (\a=e,\mu)$ is the fraction
of $\alpha$-asymmetry that is first washed-out by the $N_2$ inverse processes in the tauon-orthogonal plane and 
then by the $N_1$ inverse processes.  

The terms $\Delta p_{{\rm p}e}$, $\D p_{{\rm p}\m}$ and $\Delta p_{{\rm p}\tau}$,
with  $\D p_{{\rm p} e}+ \D p_{{\rm p}\mu} + \D p_{{\rm p}\tau} = 0$,
take into account the possibility of different flavour compositions of the pre-existing leptons and anti-leptons.
This would lead to initial values of the  pre-existing $\a$ asymmetries that are not necessarily 
just a fraction of $N^{\rm p,i}_{B-L}$. The presence of these terms depends on the specific 
mechanism that produced the pre-existing asymmetry. For example in leptogenesis itself they are 
in general present, they are the so called phantom terms. 
However,  
this indefiniteness has just a very small effect on the results.   If the $\D p_{{\rm p}\a}$-terms are not present, then 
in principle very special flavour configurations with $1-p^0_{e \t_2^{\bot}}, 1-p^0_{\mu \t_2^{\bot}} \ll 1$
could also lead to a wash-out of the pre-existing asymmetries without the need to impose
$K_{1e}, K_{1\mu} \gg 1$. We will comment on this possibility but for the time being we will assume that 
these terms are present. In this case 
the condition of successful strong thermal leptogenesis translates into the 
straightforward set of conditions 
\be\label{stc}
K_{1e}, K_{1\mu} \gtrsim  K_{\rm st}(N^{\rm p,i}_{\D_{e,\m}}), \,\, 
K_{2\t} \gtrsim K_{\rm st}(N^{\rm p,i}_{\D_{\t}}), \; K_{1\tau}\lesssim 1    \,  .
\ee  
 These conditions guarantee a washout of the electron and muon asymmetries,
only possible in the three-flavoured regime at $T \ll 10^{9}\,$GeV, and at the same time also a wash-out of the
tauon asymmetry in the two-flavoured regime. The latter is still compatible with a 
generation of a sizeable tauon asymmetry from $N_2$ decays. This is the only possibility \cite{problem}. 
It should be noticed that in the $N_2$-dominated scenario the existence of the heaviest RH neutrino $N_3$
is necessary in order to have an interference of tree level $N_2$ decays with one-loop $N_2$ decay graphs 
containing virtual $N_3$  yielding sufficiently large $\ve_{2\a}$.  
Therefore, within the $N_2$-dominated scenario, where by definition 
$M_1 \ll 10^9\,{\rm GeV}$,  one has  a phenomenological reason to have at least 
three RH neutrino species \cite{geometry}. 
\footnote{
In the limit $M_3 \ra \infty$,  when $N_3$ decouples and a two RH neutrino scenario is effectively recovered with $m_1 = 0$,
one has $\b_{2\a} \ra 0$ (cf. eq.~(\ref{eps2a})). In this limit the only possibility to realise successful leptogenesis 
is to have sizeable $C\!P$ asymmetries from the interference terms with the lightest RH neutrinos that we neglected
when we wrote eq.~(\ref{eps2a}). 
These terms are $\propto M_1$ and 
successful leptogenesis necessarily requires in the end a lower bound 
$M_1 \gtrsim 2 \times 10^{10}\,{\rm GeV}$ \cite{2RHN}. However, then in this case  
the $N_1$-produced asymmetry  not only cannot be neglected but typically dominates on the $N_2$-produced asymmetry 
and moreover, more importantly for us, strong thermal leptogenesis cannot be realised \cite{problem}. 
This well illustrates that in the $N_2$-dominated scenario,
the presence of a (coupled) $N_3$ is necessary  for successful leptogenesis.}

\subsection{Case $M_3 \lesssim 5 \times 10^{11}\,{\rm GeV}$}

If $M_3 \lesssim 5\times 10^{11}\,{\rm GeV}$, then the heaviest RH neutrinos $N_3$ 
can contribute to wash-out the tauon component 
together with the next-to-lightest RH neutrinos $N_2$.
 In this way, for the relic value of the pre-existing asymmetry, one obtains  ($\a=e,\m$)
\bea\label{finalpaslowM3}
N_{\D_\t}^{\rm p,f} & = & 
(p^0_{{\rm p}\t}+\D p_{{\rm p}\tau})\,  
e^{-{3\pi\over 8}\,(K_{1\t}+K_{2\t}+K_{3\t})} \, N_{B-L}^{\rm p,i} \,  , \\  \nonumber
N_{\D_\a}^{\rm p,f} & = &  \left\{(1-p^0_{{\rm p}\t})\,\left[p^0_{p\t_3^{\bot}}\,p^0_{\t_3^{\bot}\t_2^{\bot}}\,
p^0_{\t_2^{\bot}\a}\,e^{-{3\pi\over 8}\,(K_{3\t^{\bot}}+K_{2\tau^{\bot}})}  \right.\right.  
  +   (1-p^0_{{\rm p}\t_3^{\bot}})\,(1-p^0_{\t_3^{\bot}\t_2^{\bot}})\,p^0_{\t_2^{\bot}\a}
 \, e^{-{3\pi\over 8}\,K_{2\tau^{\bot}}} \\ \nonumber
 & + &  \left.\left. p^0_{{\rm p}\t_3^{\bot}}\,
 (1-p^0_{{\rm p}\t_2^{\bot}})\,(1-p^0_{\t_2^{\bot}\a}) \right] + \D p_{{\rm p} \a}\right\}
 \, e^{-{3\pi\over 8}\,K_{1\a}}\, N_{B-L}^{\rm p , i} \,  ,
\eea
where we defined $K_{2\t^{\bot}}\equiv K_{2e} + K_{2\mu}$ 
and $K_{3\t^{\bot}}\equiv K_{3e}+K_{3\mu}$.
The inclusion of the $N_3$-washout relaxes the condition $K_{2\t} \gg 1$ to $K_{2\tau}+K_{3\tau}\gg 1$.
In this way  one can have strong thermal leptogenesis with lower values of $K_{2\tau}$ and so the condition of 
successful leptogenesis can be more easily satisfied. Therefore, in this case the constraints from 
successful strong thermal leptogenesis could potentially get relaxed.

\section{Lower bound on neutrino masses}

In this Section we show finally that the strong thermal condition implies,
for sufficiently large pre-existing asymmetries and barring 
fine tuned conditions on the values of the flavour decay parameters and in the seesaw formula, 
the existence of a lower bound on the lightest neutrino mass and, more generally, a strong
reduction of the accessible region of parameters for $m_1 \lesssim 10\,{\rm meV}$. 

The main point is that the conditions $K_{1\tau}\lesssim 1$ and 
$K_{1e}, K_{1\mu}\gtrsim K_{\rm st} \gg  1$  can be satisfied simultaneously only for sufficiently large
values of $m_1$.  

\subsection{Case $M_3 \gtrsim 5\times 10^{11}\,{\rm GeV}$}

Let us start discussing the more significant case  $M_3 \gtrsim 5\times 10^{11}\,{\rm GeV}$,
when, as already pointed out, the $N_3$ wash-out can be neglected.
The cases of NO and IO need also to be discussed separately. Let us start from NO. 

\subsubsection{NO neutrino masses}

 We want to show that the conditions $K_{1\tau}\lesssim 1$
 and $K_{1e}, K_{1\m}\gtrsim K_{\rm st} \gg 1$ can be satisfied simultaneously,
 without fine-tuned conditions,  only if $m_1$ is sufficiently large.
Let us start by analysing $K_{1\tau}$. The general eq.~(\ref{fdp}) for the $K_{i\alpha}$'s specialises into
\be\label{k1tau}
K_{1\tau}  =  \left|\sqrt{m_1\over m_{\star}}\,U_{\t 1}\,\O_{11} 
	    + \sqrt{m_2\over m_{\star}}\,U_{\t 2}\,\O_{21}  
	    +  \sqrt{m_3\over m_{\star}}\,U_{\t 3}\,\O_{31} \right|^2  \,  .
\ee
From this expression, anticipating that the lower bound falls into a range of values $m_1 \lesssim m_{\rm sol}$ so
that we can approximate $m_2 \simeq m_{\rm sol}$ and $m_3 \simeq m_{\rm atm}$, we can
write 
\be\label{condition}
\sqrt{m_{\rm atm}\over m_{\star}}\,U_{\t 3}\,\O_{31} \simeq 
-  \sqrt{m_{1}\over m_{\star}}\, U_{\t 1} \, \O_{11} 
-  \sqrt{m_{\rm sol}\over m_{\star}}\,U_{\t 2} \, \O_{21} 
+ \sqrt{K_{1\t}} \, e^{i\varphi}   ,
\ee
where $\varphi$ is some generic phase. 
If we now insert this expression into the expressions for $K_{1e}$ and $K_{1\mu}$,
we can impose ($\a = e, \m$)
\be
K_{1\a} \simeq \left|\O_{11} \, \sqrt{m_{1}\over m_{\star}} \, \left(U_{\a 1} 
- {U_{\t 1}\over U_{\t 3}} \, U_{\a 3} \right) + \sqrt{K_{1\a}^0}\,e^{i\,\varphi_0} \right|^2
> K_{\rm st}(N^{\rm p,i}_{\D\a}) \,  ,
\ee
where we defined $K_{1\a}^0 \equiv K_{1\a}(m_1=0)$ and $\varphi_0$ such that
\be
\sqrt{K_{1\a}^0} \, e^{i\,\varphi_0}  \equiv \O_{21} \, \sqrt{m_{\rm sol}\over m_{\star}} \, \left(U_{\a 2} - {U_{\t 2}\over U_{\t 3}} \, U_{\a 3} \right) 
+ {U_{\a 3}\over U_{\t 3}} \,\sqrt{K_{1\t}}\,e^{i\,\varphi}  \,  .
\ee
 From this condition one obtains a lower bound on $m_1$  ($\a=e,\m$),
\be\label{lb}
m_1 > m_1^{\rm lb} \equiv
m_{\star} \,  
{\rm max}_\a\left[
\left({\sqrt{K_{\rm st}}- \sqrt{K^{0,{\rm max}}_{1\a} }\over 
{\rm max}[|\O_{11}|]\left|U_{\a 1} - {U_{\t 1}\over U_{\t 3}} \, U_{\a 3} \right| }\right)^2 
\right] \,  
\ee
when $K^{0,{\rm max}}_{1\a} < K_{\rm st}$, where we defined
\be\label{K10max}
K^{0,{\rm max}}_{1\a} \equiv \left({\rm max}[|\O_{21}|]\,\sqrt{m_{\rm sol}\over m_{\star}} \, 
\left|U_{\a 2} - {U_{\t 2}\over U_{\t 3}} \, U_{\a 3} \right| + 
\left|{U_{\a 3}\over U_{\t 3}} \right|\,\sqrt{K_{1\tau}^{\rm max}} \right)^2 \, .
\ee
Because of the smallness of the reactor mixing angle $\theta_{13}$ there are two 
consequences: the first is that the maximum is found for $\a=e$ and the second is that, imposing
$K_{1\t}^{\rm max} \lesssim 1$, both the two terms in $K^{0,{\rm max}}_{1e} $ proportional to $U_{e3}$ are suppressed 
and in this way there is indeed a lower bound  
for a sufficiently small value of ${\rm max}[|\O_{21}|]$. 

In the left panel of Fig.~1 we have conservatively taken ${\rm max}[|\O^2_{21}|] = {\rm max}[|\O^2_{11}|] = M_{\O}=2$ 
and plotted $m_1^{\rm lb}$  at $95\%\, {\rm C.L.}$ 
for $N^{\rm p}_{B-L} = 0.1$ as a function of the Dirac phase $\delta$. 
\footnote{We used Gaussian ranges for the mixing angles within as in eq.~(\ref{mixinganglesNO}),
except for the atmospheric mixing angle for which we used a Gaussian distribution 
 $s^2_{23}=0.5 \pm 0.1$, i.e. centred on the maximal mixing value since on this angle results
 are still unstable depending on the analysis. We have also used, in the scatter plot analysis as well, 
 $p^0_{{\rm p} \tau_2^{\bot}}/2= p^0_{{\rm p} \t}=\D p_{\rm p e} = \D p_{{\rm p} \m} = 1/3$,
 corresponding to a flavour blind pre-existing asymmetry. Notice in any case that 
 results depend only logarithmically on these parameters, so they are insensitive to a precise choice.} 
 \begin{figure}
\begin{center}
\hspace{-3mm}
\psfig{file=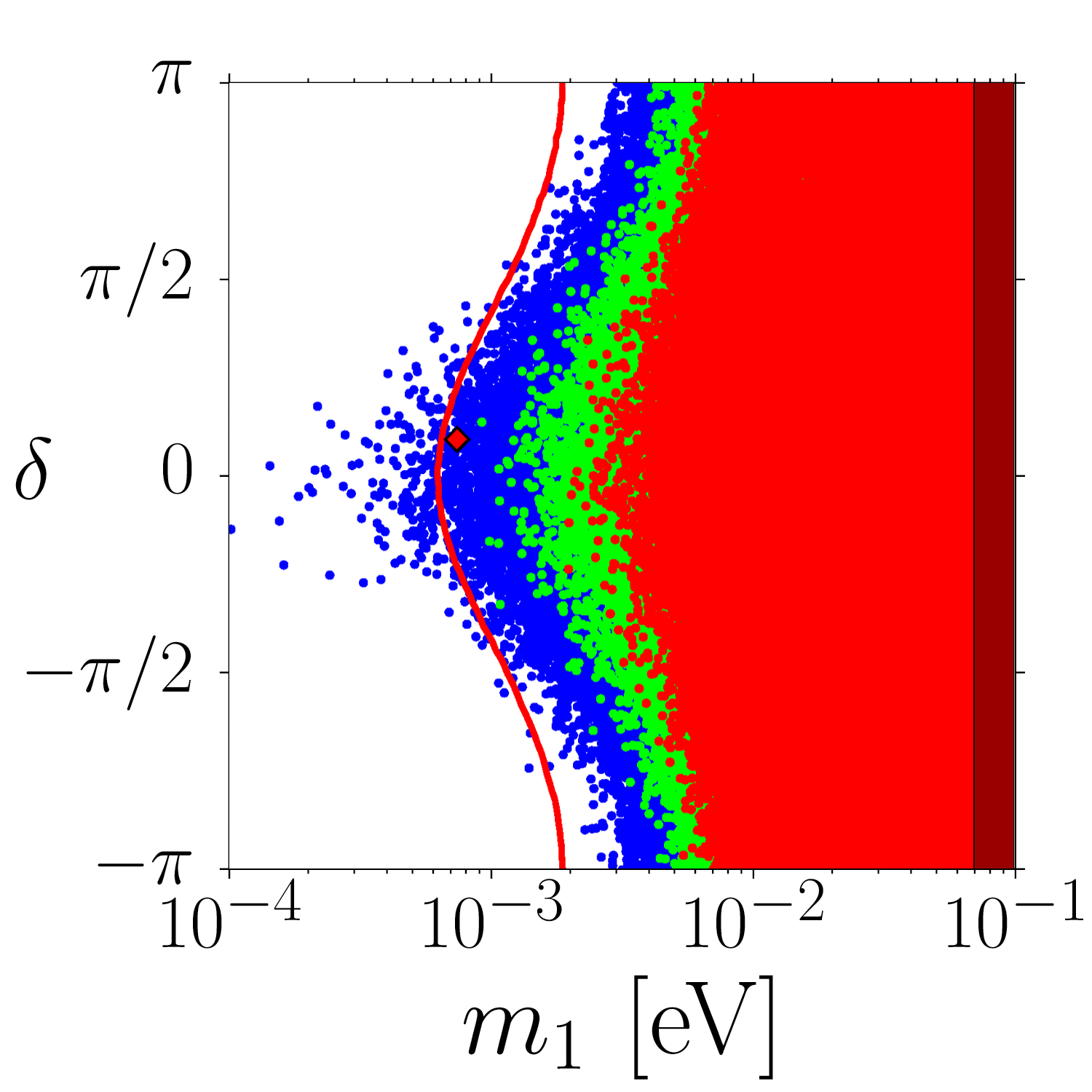,height=47mm,width=50mm}
\hspace{1mm}
\psfig{file=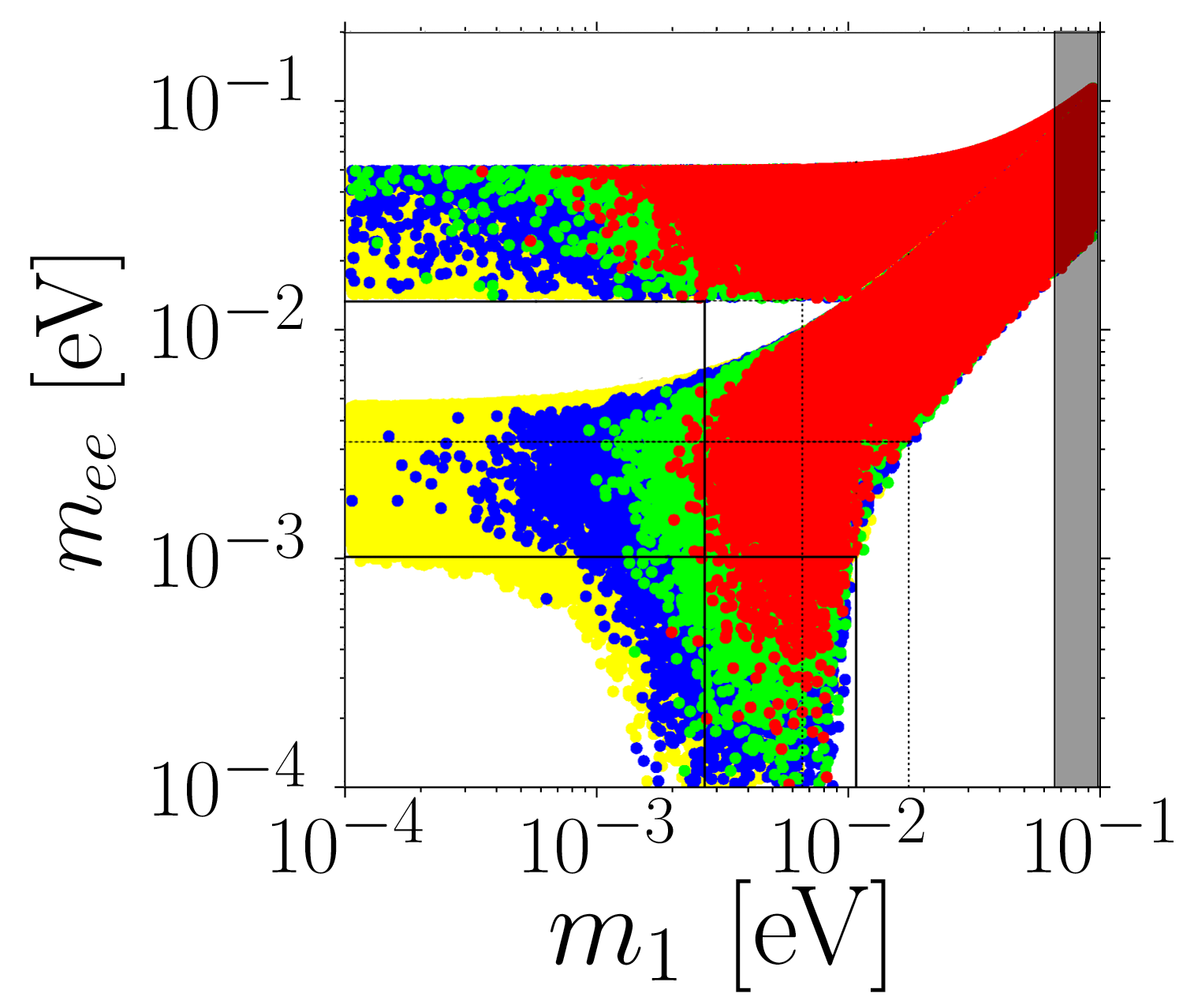,height=47mm,width=54mm} 
\hspace{1mm}
\psfig{file=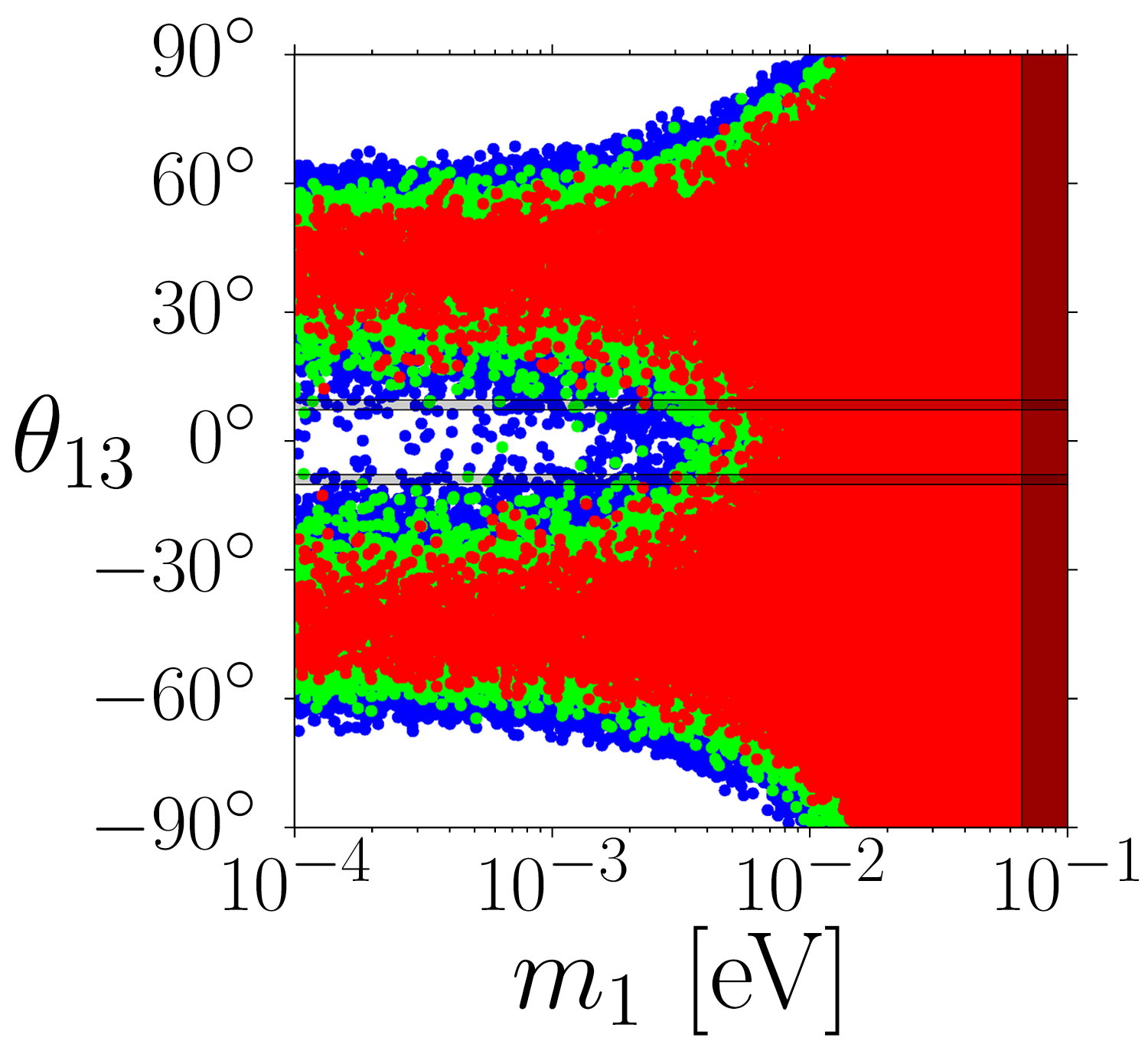,height=47mm,width=50mm}
\end{center}\vspace{-8mm}
\caption{NO case.  Scatter plot points in the planes $\d-m_1$ (left), $m_{ee}-m_1$ (center), 
$\theta_{13}-m_1$ (right) satisfying  successful strong thermal leptogenesis  
for $N_{B-L}^{\rm p, i}=10^{-1}$ (red), $10^{-2}$ (green) and $10^{-3}$ (blu). 
In all panels the vertical gray band is the {\em Planck} $m_1$ upper bound eq.~(\ref{PLANCK}).
In the left panel points are plotted  for $M_{\Omega}=2$ and  
the red solid line is the analytic lower bound $m_1^{\rm lb}(\d)$ (cf. eq.~(\ref{lb})) for $N_{B-L}^{\rm p,i,}=10^{-1}$.  
While the points in the left and central panels have been obtained for uniform random values of
the three mixing angles generated within the $3\s$ ranges eq.~(\ref{mixinganglesNO}), in the right
panel they have been left free (the horizontal band indicates the $3\,\s$ range in 
eq.~(\ref{mixinganglesNO}) for $\theta_{13}$). In the central panel the vertical lines 
indicate the $m_1$ values above which $99\%$ of scatter plot points are found  (see central panel in Fig.~4).}
\label{deltamee}
\end{figure}
 At $\d=0$ we find  (top right panel) $m_1^{\rm lb} \simeq 0.7\,{\rm meV}$ 
 while for $\d = \pm \pi$ we obtain $m_1^{\rm lb} \simeq 2\,{\rm meV}$, showing how a future determination
 of the Dirac phase $\d$ could tighten the lower bound.  
  The lower bound becomes more
 stringent for $M_{\O}=1$ and we find $m_1^{\rm lb}(\d=0) \simeq 6\,{\rm meV}$.  
 On the other hand for $M_{\Omega}=3$ the lower bound gets relaxed 
 and we obtain $m_1^{\rm lb}(\d=0) \simeq 0.13\,{\rm meV}$.
For  $M_{\O}\gtrsim 4$ one can easily verify that  the condition 
$K_{\rm st} > K^{0,{\rm max}}_{1 \a} \, (\a=e,\m)$ is not verified 
 and there is no lower bound on $m_1$.  

In order to verify the existence of the lower bound, to test the validity of the analytic
estimation and to show in more detail the level of fine tuning involved
in order to saturate the lower bound,  we performed a scatter plot analysis in the space 
of the 13 parameters ($m_1$, 6 in $U$, 6 in $\O$)
for $M_{\Omega}=1,2,5,10$. The results are shown in Fig.~1.  for three values of 
$N^{\rm p,i}_{B-L} = 10^{-1}, 10^{-2}, 10^{-3}$ (respectively the red, green and blue points). 
One can see that for $N^{\rm p,i}_{B-L} = 10^{-1}$ the minimum values of $m_1$ in the left panel at different
values of $\d$ are much higher than the analytic estimation (one has to compare the red points with the red solid line).
The reason is due to the fact that the lower bound is saturated for very special choices of $\O$ such that 
${\rm max}[|\O^2_{11}|], {\rm max}[|\O^2_{21}|]$ are as close as possible to the maximum value $M_{\O}$ 
but at the same time not to suppress too much the $C\!P$ asymmetry $\ve_{2\t}$ needed to have 
successful leptogenesis. This is confirmed by Fig.~2 where in the three panels 
we have plotted $\beta_{2\tau}\equiv\ve_{2\tau}/\bar{\ve}(M_2)$,
$|\O^2_{11}|$ and $|\O^2_{22}|$ for $M_{\O}=2$.  We have made a focused search
(by fine-tuning the parameters) managing to find a point
 (the red diamond) where $m_1$ is very close to the lower bound.
 For this point $\beta_{2\tau}$ gets considerably reduced since 
 it corresponds to a situation where the term $\propto \sqrt{m_1}$ in the flavoured decay parameters 
 becomes negligible and the strong thermal condition is satisfied for a very special condition,
 basically the eq.~(\ref{condition})  when the terms $\propto \sqrt{m_1}, \sqrt{K_{1\tau}}$ are neglected in the right-hand side and $|\O_{11}|, |\O_{21}|$ become maximal, that leads to a $C\!P$ asymmetry suppression. 
\begin{figure}
\begin{center}
\hspace{-6mm}
\psfig{file=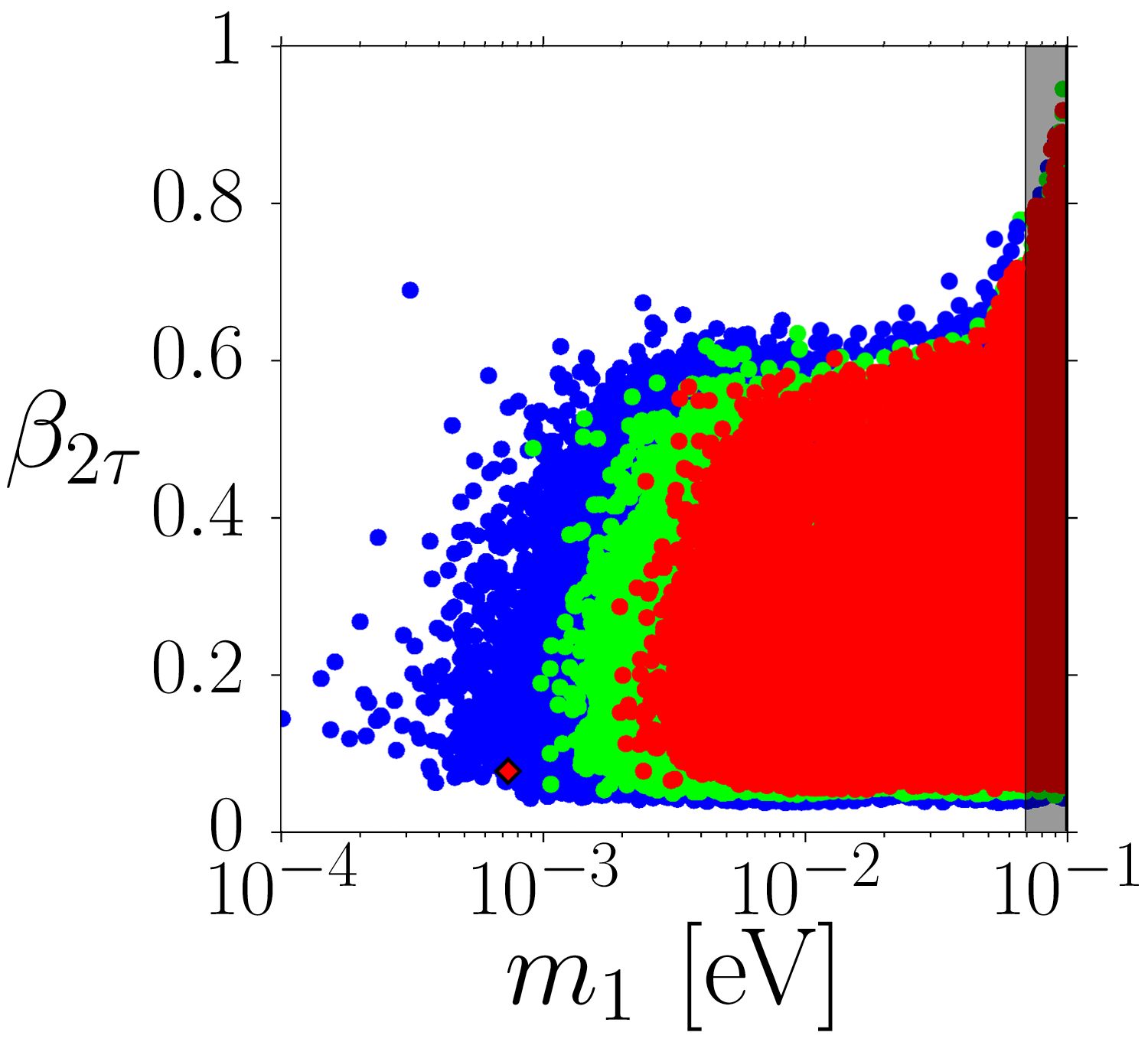,height=47mm,width=52mm} 
\psfig{file=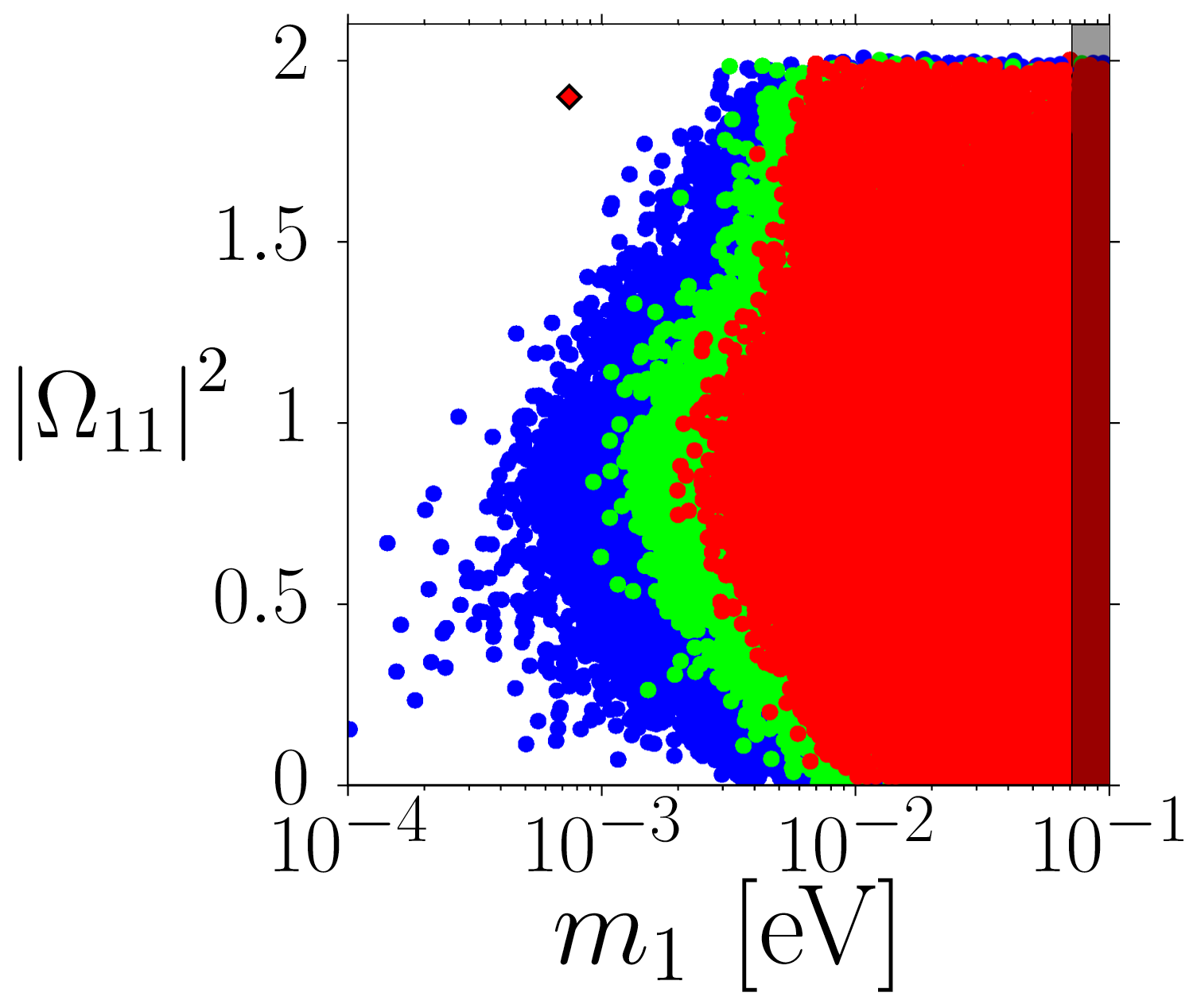,height=47mm,width=55mm}
\psfig{file=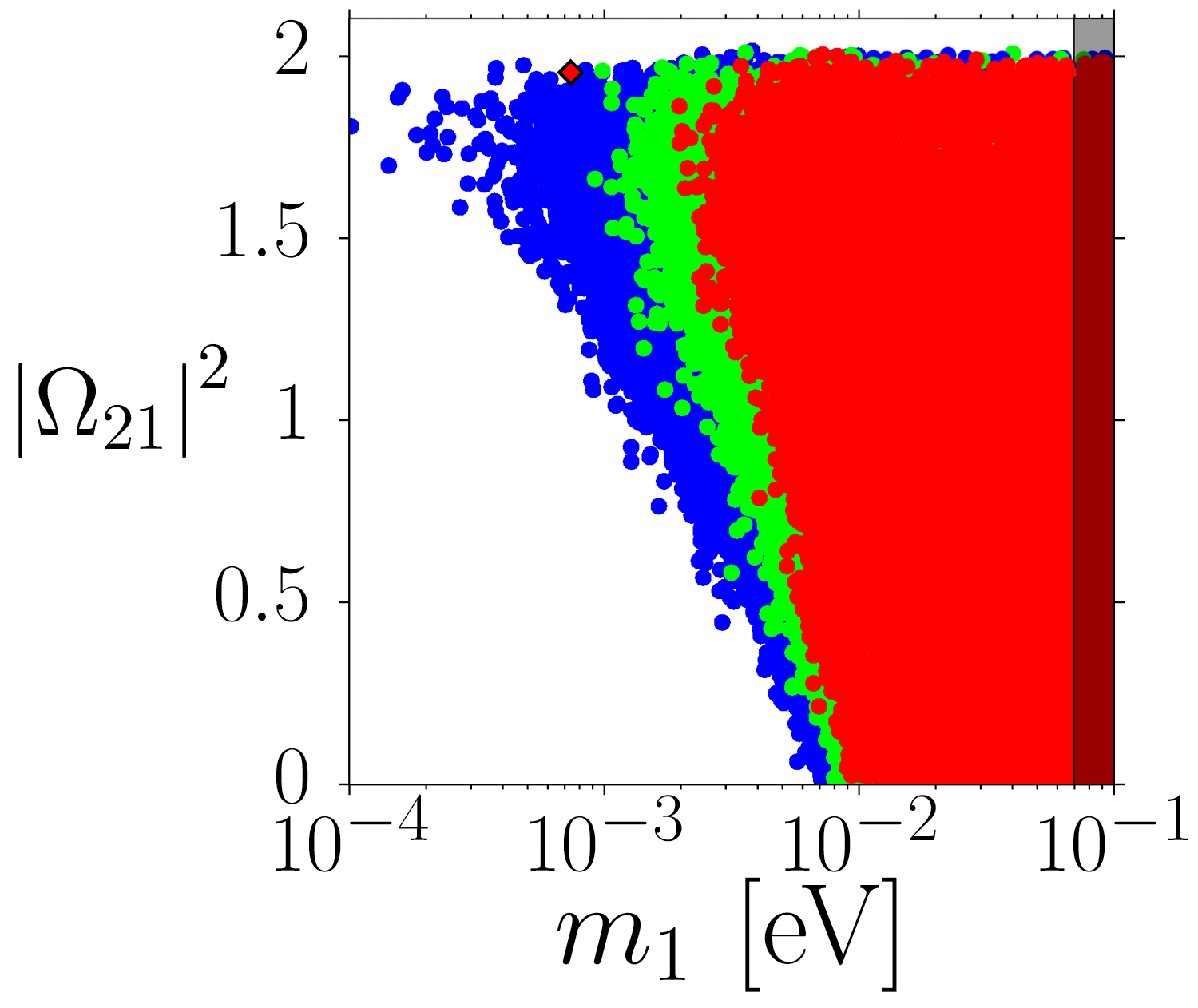,height=47mm,width=55mm}
\end{center}\vspace{-6mm}
\caption{NO case. Results of the scatter plots for $M_{\O}=2$ for $\beta_{2\tau}\equiv \ve_{2\tau}/\bar{\ve}(M_2)$,
$|\O_{11}^2|$ and $|\O_{21}^2|$ versus $m_1$ (same colour code as in Fig.~1).}
\label{densities}
\end{figure}

We have also performed a scatter plot letting the mixing angles to vary 
within the whole range of physical values with no experimental constraints. 
In the right panel of Fig.~1 we show the results in the plane $m_1-\theta_{13}$. One can see how
the smallness of $\theta_{13}$ is crucial for the existence of the lower bound. This can be well understood
analytically considering that in the expression for  $K_{1e}^{0,\rm max}$ there are 
two terms $\propto |U_{e3}|^2$  (cf. eq.~(\ref{K10max})).  

In Fig.~3 we also show the results for the values of the three $K_{1\a}$ ($\a=e,\m,\t$)
and for $K_{2\t}$, the four relevant flavoured decay parameters, for $M_{\Omega}=2$. 
First of all one can see how the values of the flavoured decay parameters respect the strong thermal 
conditions eq.~(\ref{stc}). However, the most important plot is that one for $K_{1e}$, showing how
for values $m_1 \lesssim 10\,{\rm meV}$ the maximum value of $K_{1e}$ gets considerably reduced
until it falls below $K_{\rm st}$, indicated by the horizontal dashed line for $N^{\rm p,i}_{B-L}=0.1$, 
at the $m_1$ lower bound value (very closely realised by the red diamond point). 
It is also clear that already below $\sim 10\,{\rm meV}$
the possibility to realise strong thermal leptogenesis requires a high fine tuning in the parameters
since in  this case $K_{1e} \lesssim K_{1e}^{0,{\rm max}} \simeq 4\,M_{\O} \lesssim K_{\rm st}$ for
large asymmetries and not too unreasonably high values of $M_{\O}$.
\begin{figure}
\begin{center}
\hspace{-6mm}
\psfig{file=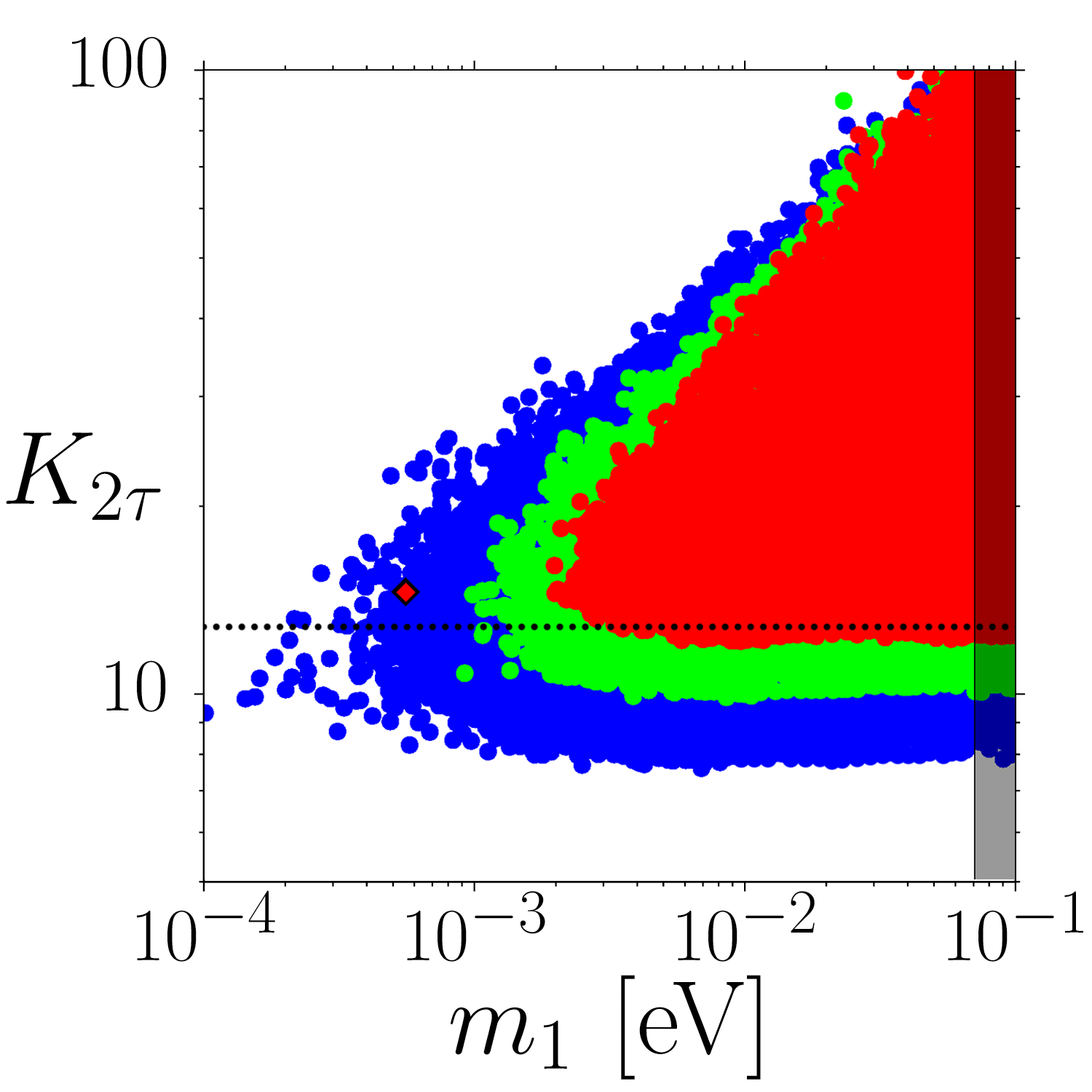,height=37mm,width=39mm}
\psfig{file=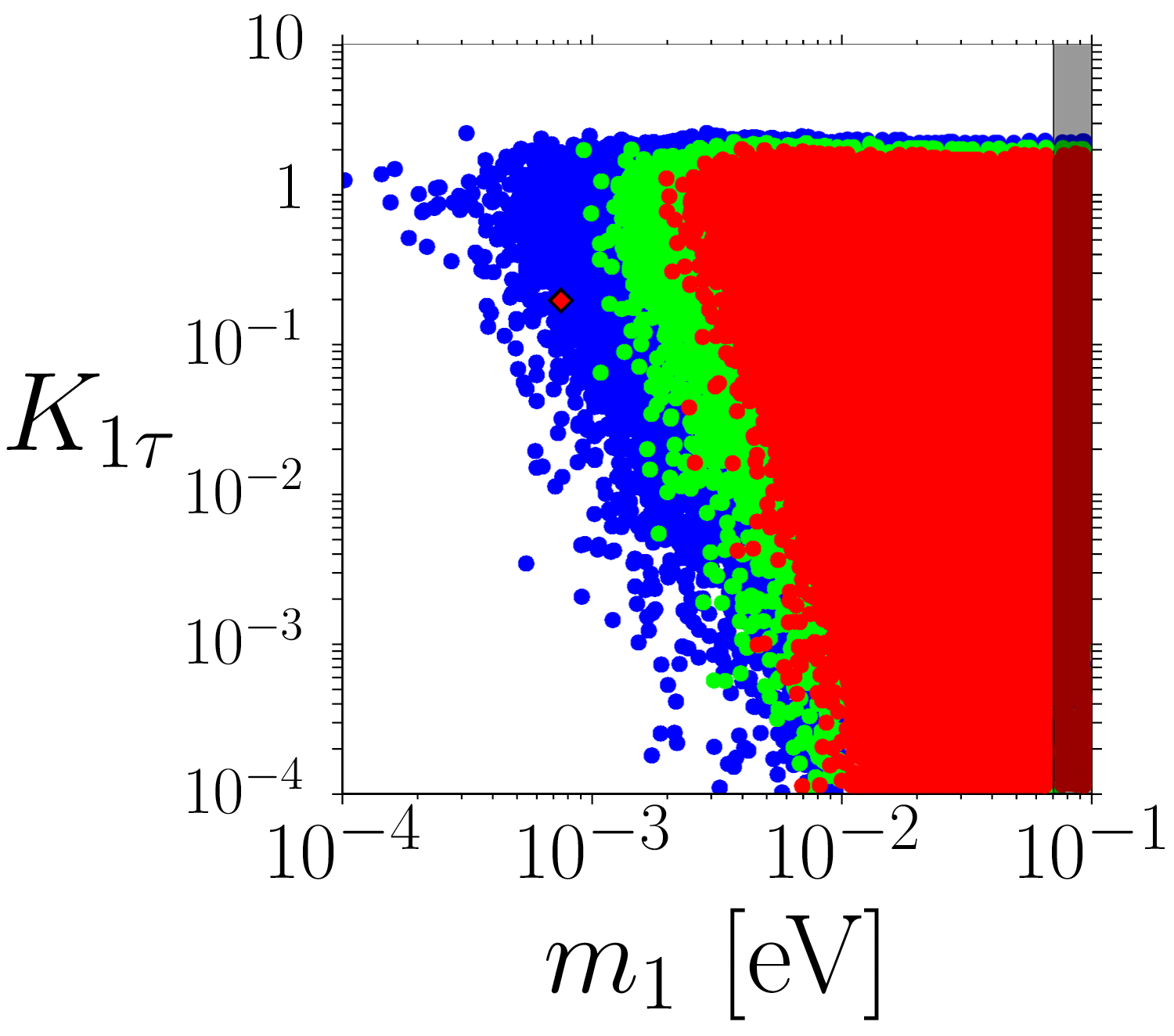,height=37mm,width=44mm}   
\psfig{file=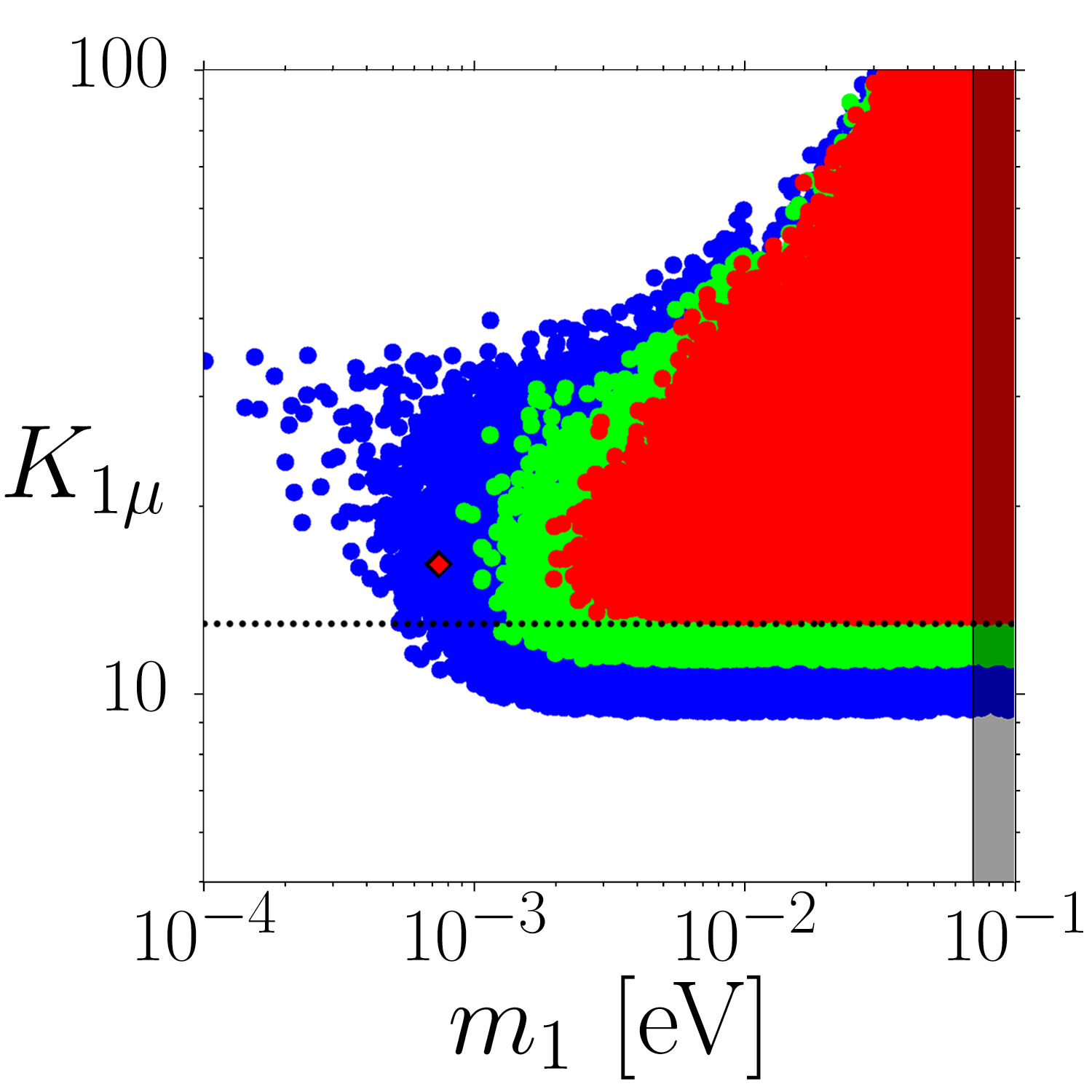,height=37mm,width=39mm}
\psfig{file=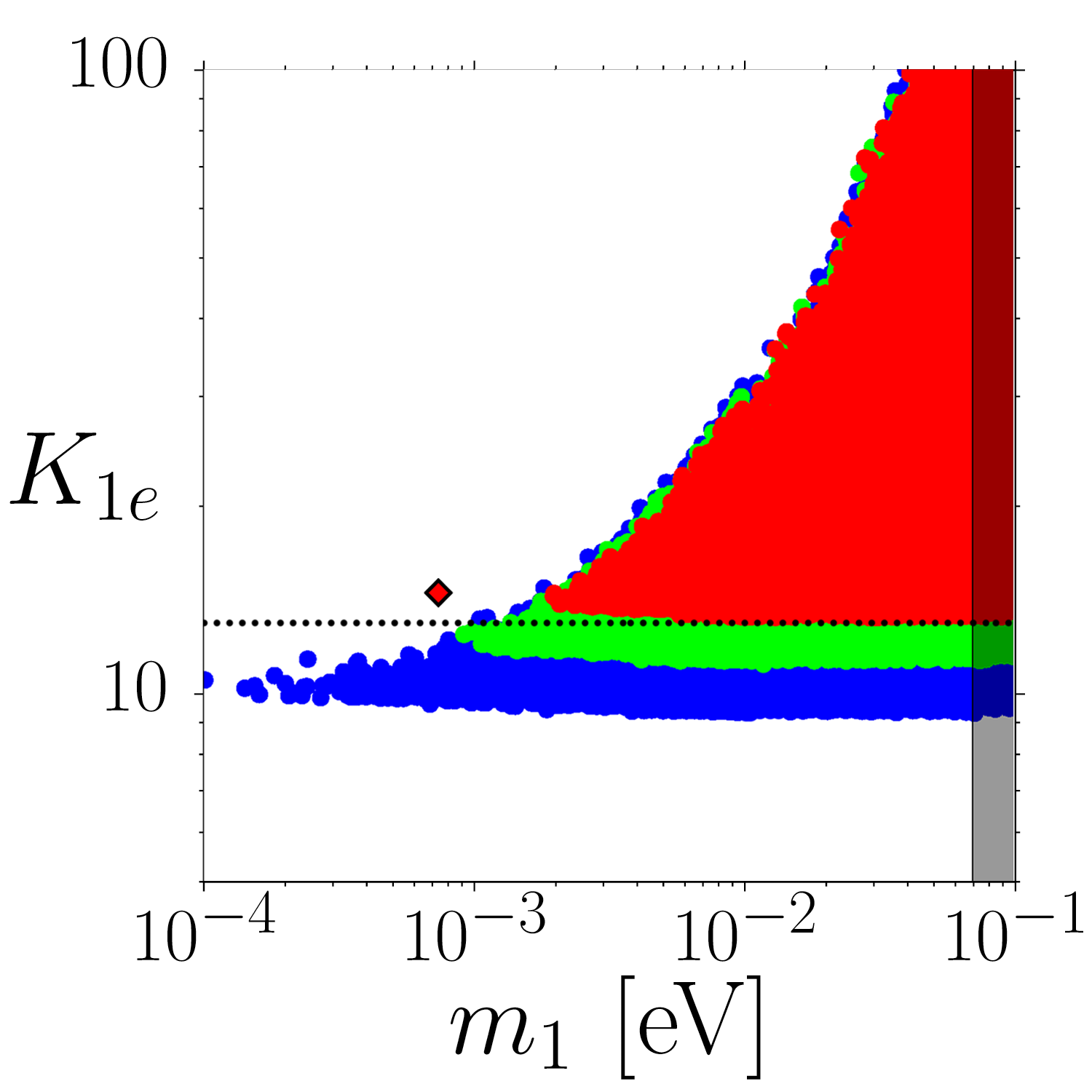,height=37mm,width=39mm}
\end{center}\vspace{-10mm}
\caption{NO case. Results of the scatter plots for $M_{\Omega}=2$ for the four relevant flavoured
decay parameters $K_{2\t}, K_{1\t}, K_{1\mu}, K_{1e}$ versus $m_1$ (conventions as in Fig.~1). The horizontal dashed
line indicates the value $K_{\rm st}(N^{\rm p,i}_{\D_\a}=0.03)\simeq 13$ (cf. eq.~(\ref{Kstrong})).}
\label{fdparameters}
\end{figure}
This is well illustrated in Fig.~4 where we plotted the distribution of the  $m_1$ values  from the scatter plots
for  $M_{\O}=1,2,5,10$ and for $N_{B-L}^{\rm p,i}= 10^{-1}, 10^{-2}, 10^{-3}$. 

\begin{figure}
\begin{center}
\vspace{-3mm}
\psfig{file=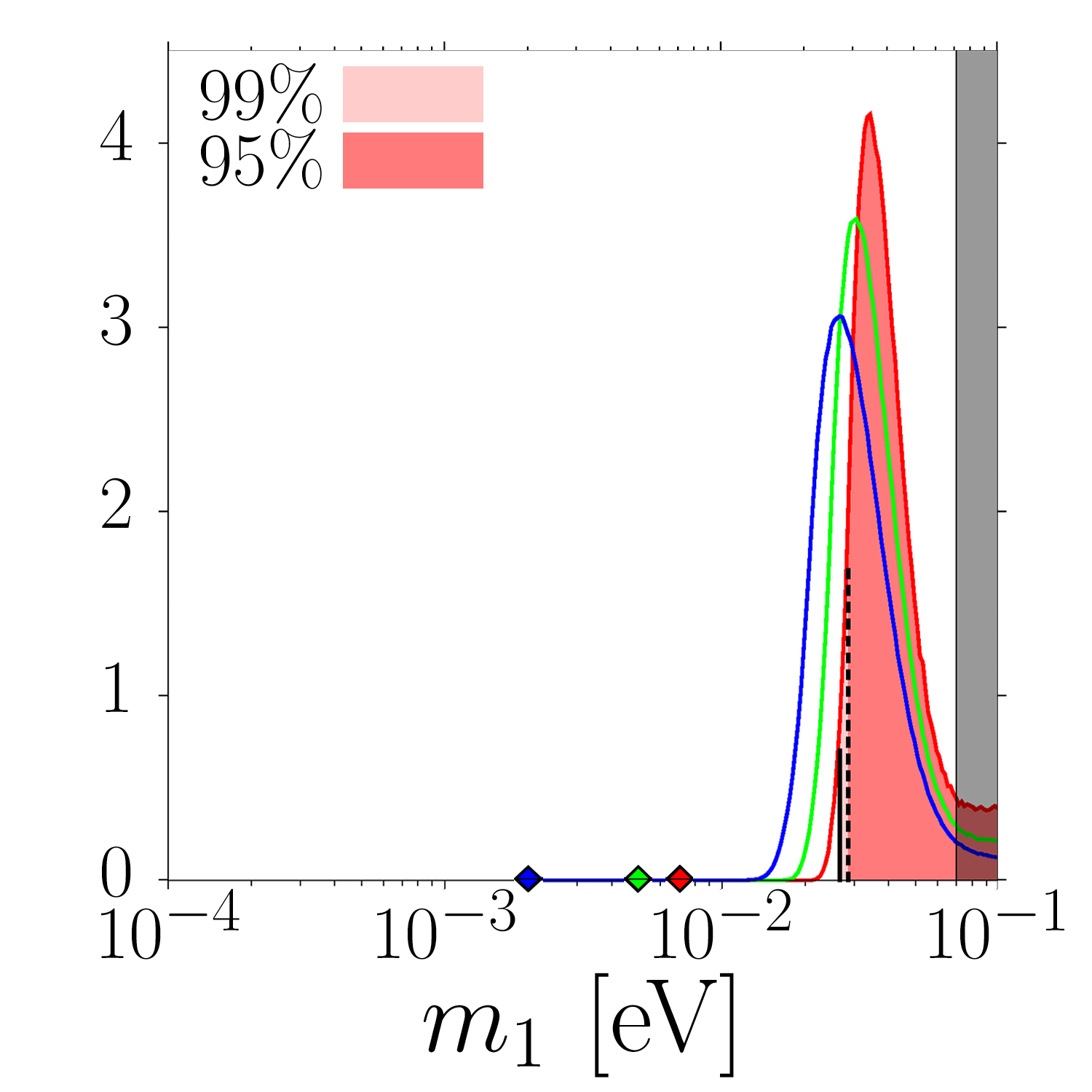,height=37mm,width=39mm}
\psfig{file=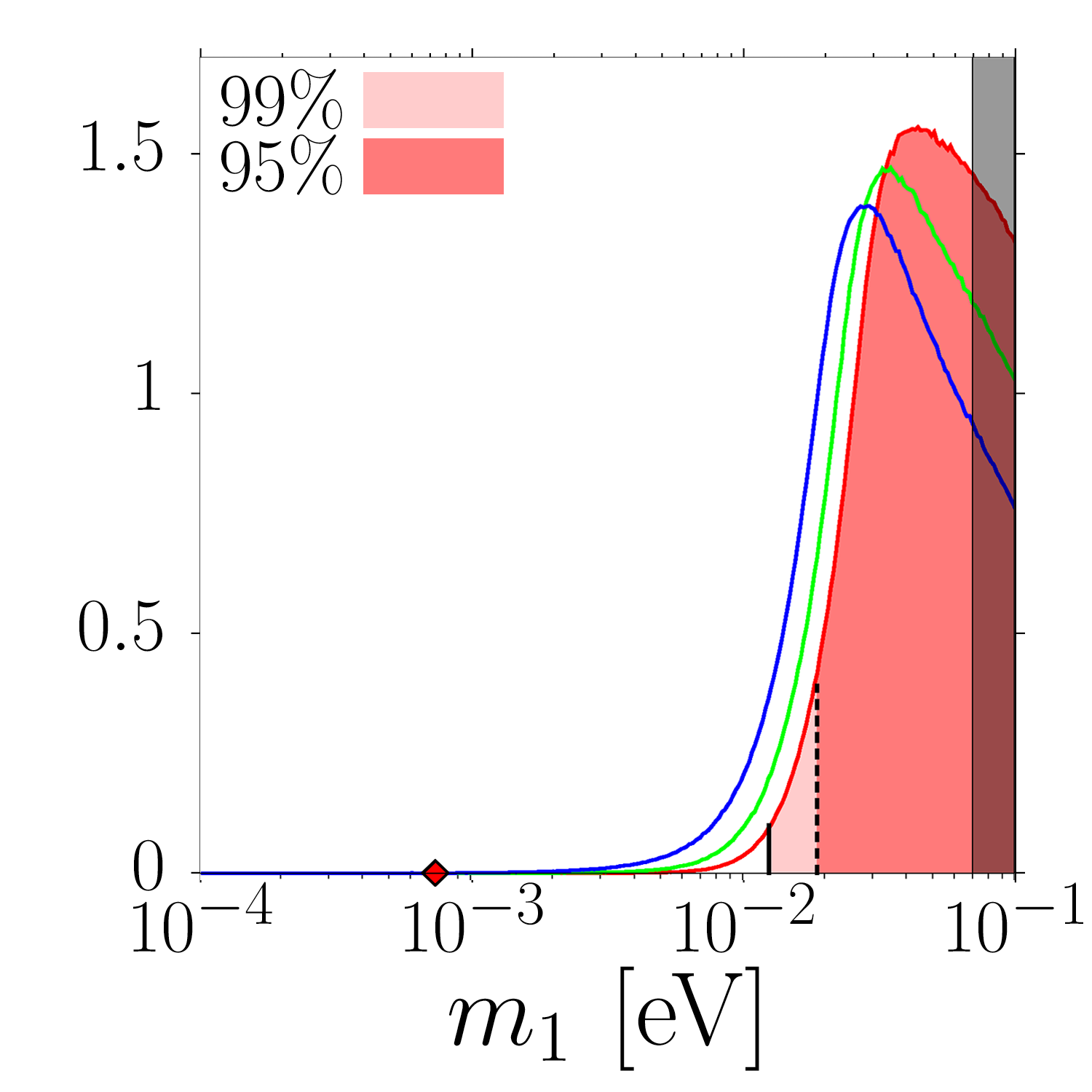,height=37mm,width=39mm} 
\psfig{file=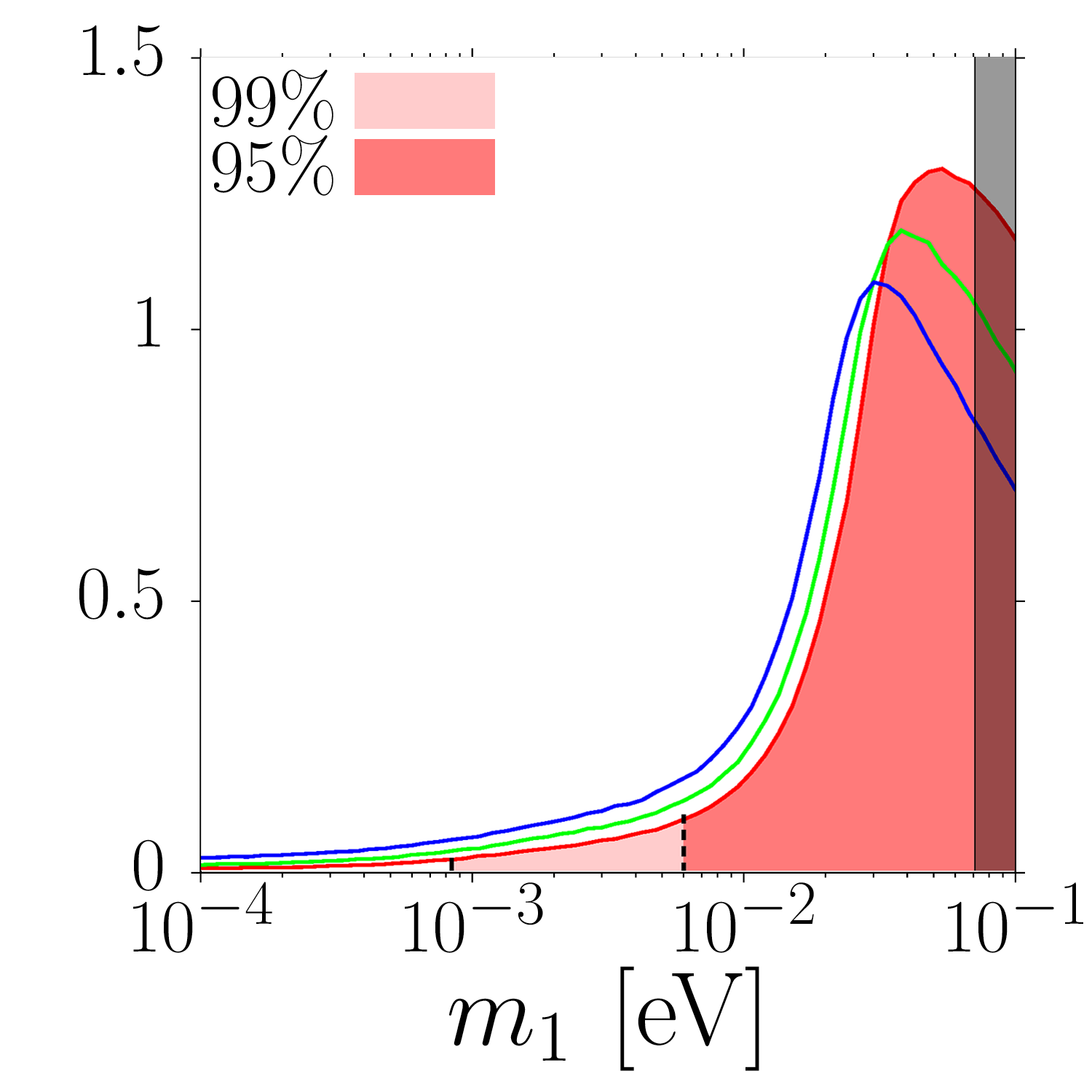,height=37mm,width=39mm}
\psfig{file=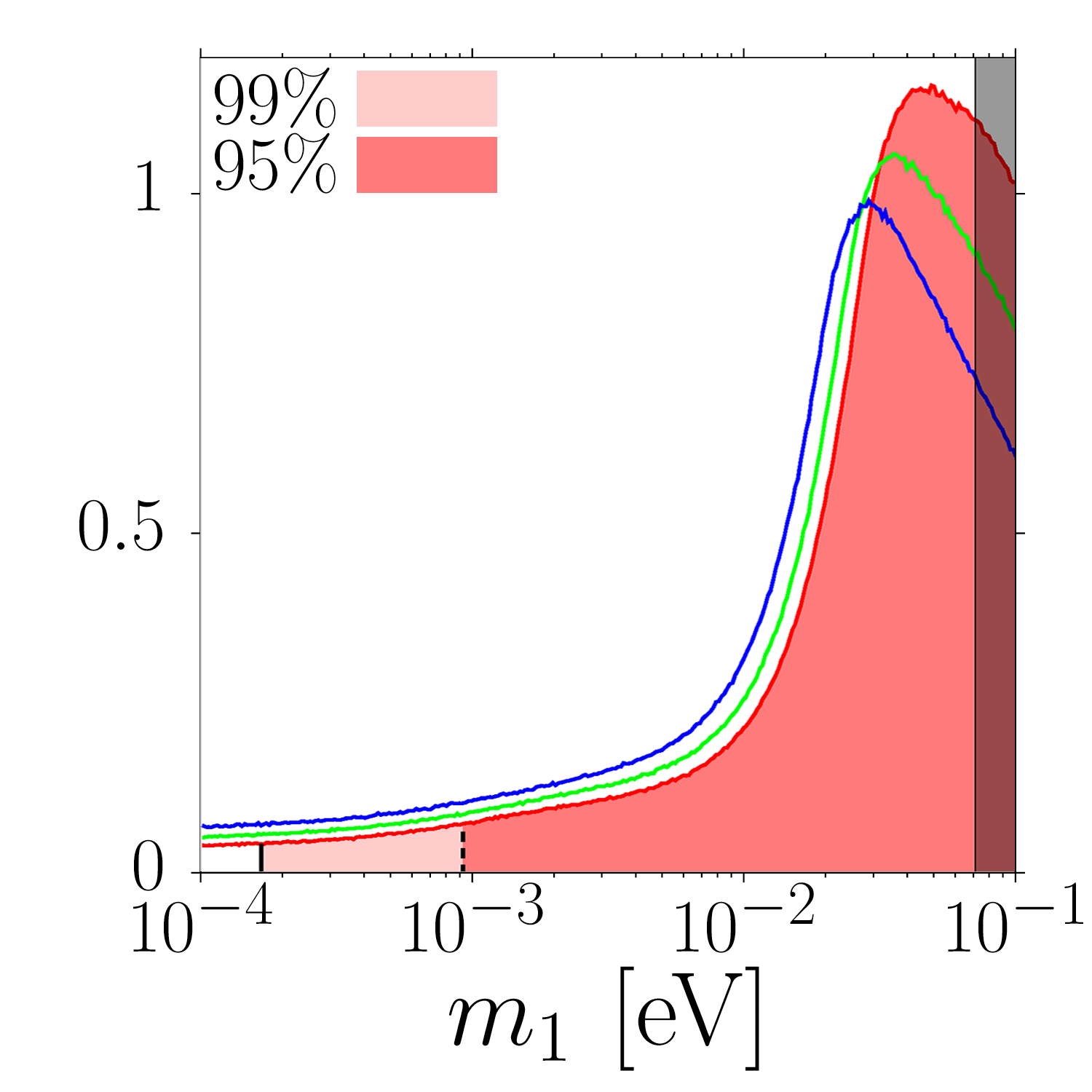,height=37mm,width=39mm}
\end{center}\vspace{-10mm}
\caption{NO case. Distribution of probability of $m_1$ from the scatter plots for $M_{\O}=1,2,5,10$
from left to right for different values of $N_{B-L}^{\rm p,i}$ (same conventions as in Fig.~1). The diamonds mark the 
$m_1$ minimum value (if found).}
\label{densities}
\end{figure}
One can see that there is a clear peak around $m_1 \simeq m_{\rm atm}$. 
One can also see that the distributions rapidly tend to zero when $m_1 \lesssim m_{\rm sol} \simeq 10\,{\rm meV}$.
For example, for our benchmark value $M_{\O}=2$ and for $N_{B-L}^{\rm p,i}= 10^{-1}$, 
it can be noticed how more than
$99 \%$ of points falls for values $m_1 \gtrsim 10\,{\rm meV}$ (the value quoted in the abstract).
Even for $M_{\Omega}=5$ one still has that the $95\%$ of points satisfying successful  strong thermal leptogenesis
is found for $m_1 \gtrsim 6\,{\rm meV}$. It is also interesting to notice how this constraint gets only slightly relaxed
for lower values of the pre-existing asymmetry.  Only for $M_{\O}=10$ one obtains that $95\%$ of points fall
at $m_1 \gtrsim 1\,{\rm meV}$. For $M_{\O}=100$, not shown in the plots, this would decrease at (untestable)
values $m_1 \gtrsim 0.4\,{\rm meV}$.
This provides another example of how, more generally, leptogenesis neutrino mass bounds tend to disappear in the
limit $M_{\O} \gg 1$ \cite{hambye}. It should be however said how large values of $|\O^2_{ij}|$ imply high cancellations
in the see-saw formula such that the lightness of LH neutrinos becomes a combined effect of these cancellations with the
the see-saw mechanism and they are typically not realised in models embedding a genuine minimal type I see-saw mechanism.

Clearly the results on the $m_1$ distributions in Fig.~4  depend on the orthogonal matrix parameterisation that
we used in order to generate the points on the scatter plots but they provide quite a useful indication 
of the level of fine tuning required to satisfy successful strong thermal leptogenesis 
for values of the lightest neutrino mass below  $\sim 10\,{\rm meV}$.  In any case it is fully 
explained by our analytical discussion and 
by the plot of the  maximum of $K_{1e}$ values that is independent of the specific parameterisation. 
 We also double checked the results producing scatter plots for two different parameterisations. 
In a first case we used the usual parameterisation of
 the orthogonal matrix in terms of complex rotations described by three complex Euler angles, 
 that, however, has the drawback not to be flavour blind. In a second case  
 we used a parameterisation based on the  isomorphism between the group of complex orthogonal 
 matrices  and the  Lorentz group. We did not find any appreciable difference. 
 \footnote{As a technical detail it is probably worth to stress that 
 for the first time we have randomly generated complex orthogonal matrices 
 (about 10 million of points for both parameterisations)  within the whole 6-dim parameter space,  without any restriction 
 (except for the bound $|\O^2_{ij}|<M_{\O}$).}   

\subsubsection{IO neutrino masses}

Let us now discuss the case of IO.  The analytical procedure we have discussed 
for NO can be repeated in the IO case and one finds the same expression eq.~(\ref{lb})
for the lower bound on $m_1$ where, however, one has to replace 
$m_{\rm sol}\ra m_{\rm atm}$ and $U \ra U^{(IO)}$. 

The replacement $m_{\rm sol} \ra m_{\rm atm}$ tends to push all $K_{1\a}$ values
to much higher values and this is indeed what happens for $K_{1e}$.  If one considers
again the quantity $K_{1e}^{0,{\rm max}}$ (cf. eq.~(\ref{K10max})) it is possible to check that this time
one has always $K_{1e}^{0,\rm max} \gg K_{\rm st}$ for $N^{\rm p, i}_{B-L} \lesssim 0.1$. On the other hand this time
the value of $K_{1\m}$ has to be fine tuned in order to be greater than $K_{\rm st}$.
The reason is that for IO there is now a cancellation in the quantity
$[U_{\mu 2}-U_{\tau 2}\,U_{\mu 3}/U_{\tau 3}]^{(IO)}$ that suppresses 
$K_{1\m}^{0,{\rm max}}$ though not as strongly as $K_{1 e}^{0,{\rm max}}$ in the NO case. 
Indeed one finds now that $K_{1\m}^{0,{\rm max}} < K_{\rm st}$, the condition for the existence 
of the lower bound, holds only for $M_{\O} \lesssim 0.9$.  This implies that 
the lower bound on $m_1$ for IO is much looser than for the NO case. 
This result is again confirmed by a scatter plot analysis. The results are shown
in Fig.~5 directly in the form of the distribution of probabilities for $m_1$. 
 \begin{figure}
\begin{center}
\psfig{file=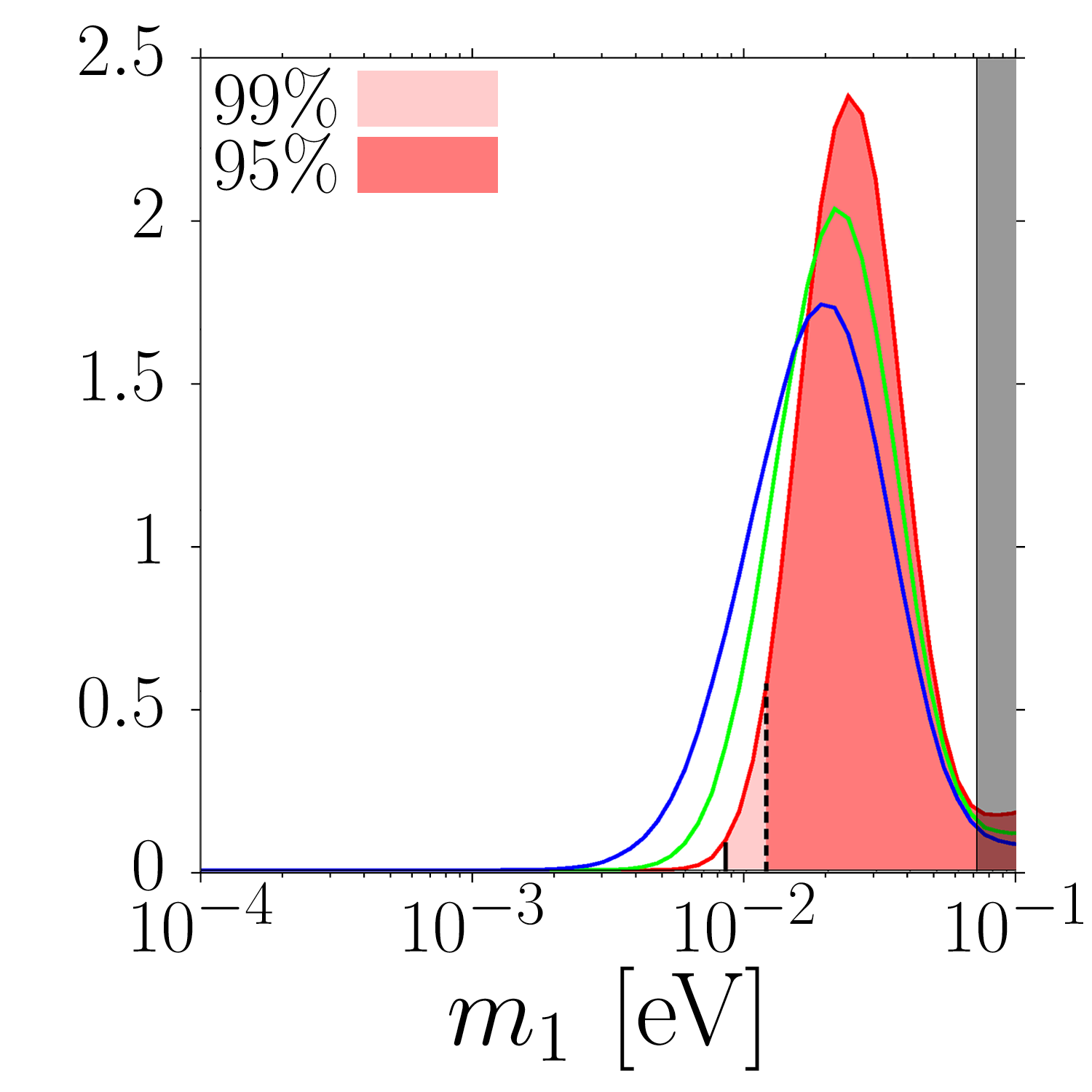,height=37mm,width=39mm}
\psfig{file=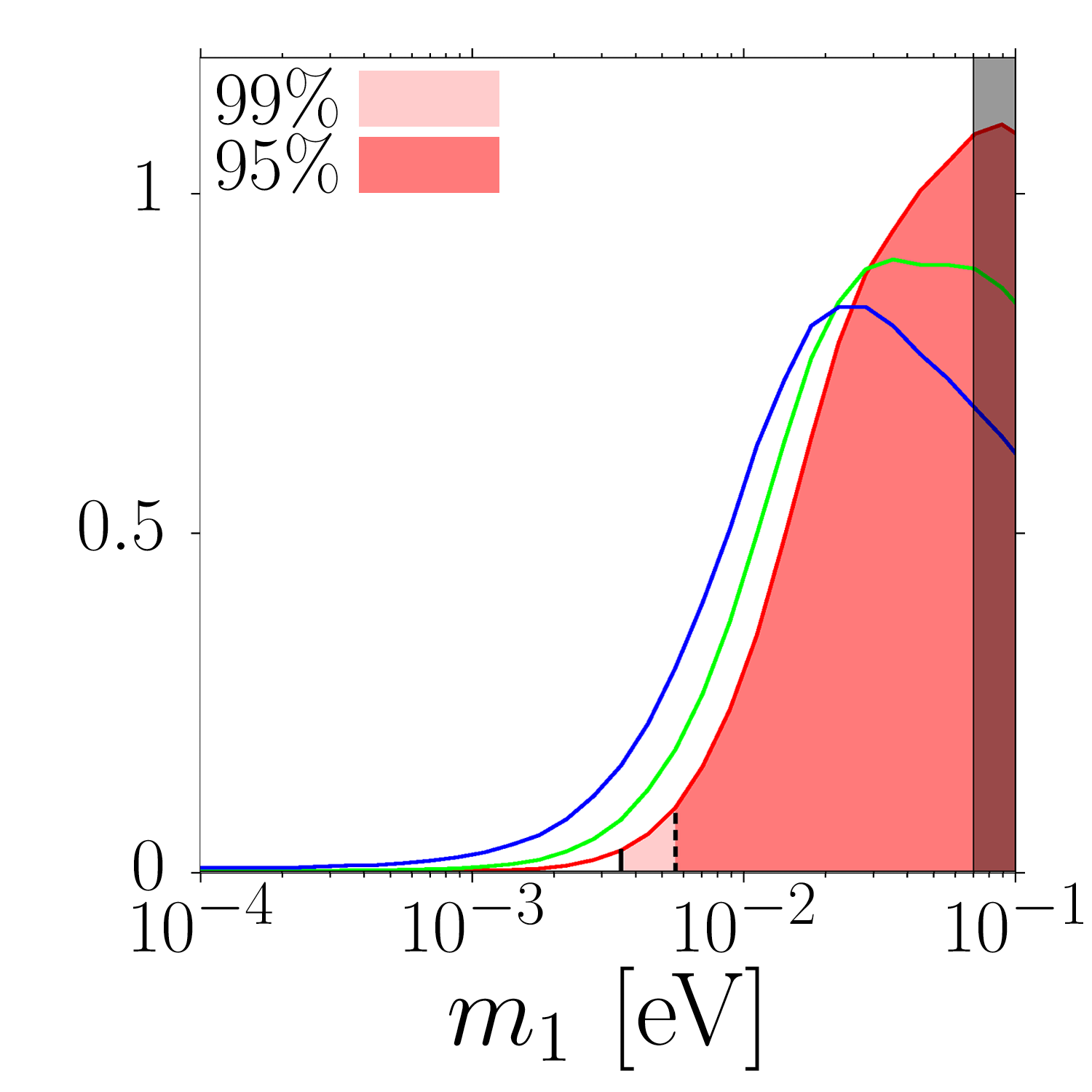,height=37mm,width=39mm}  
\psfig{file=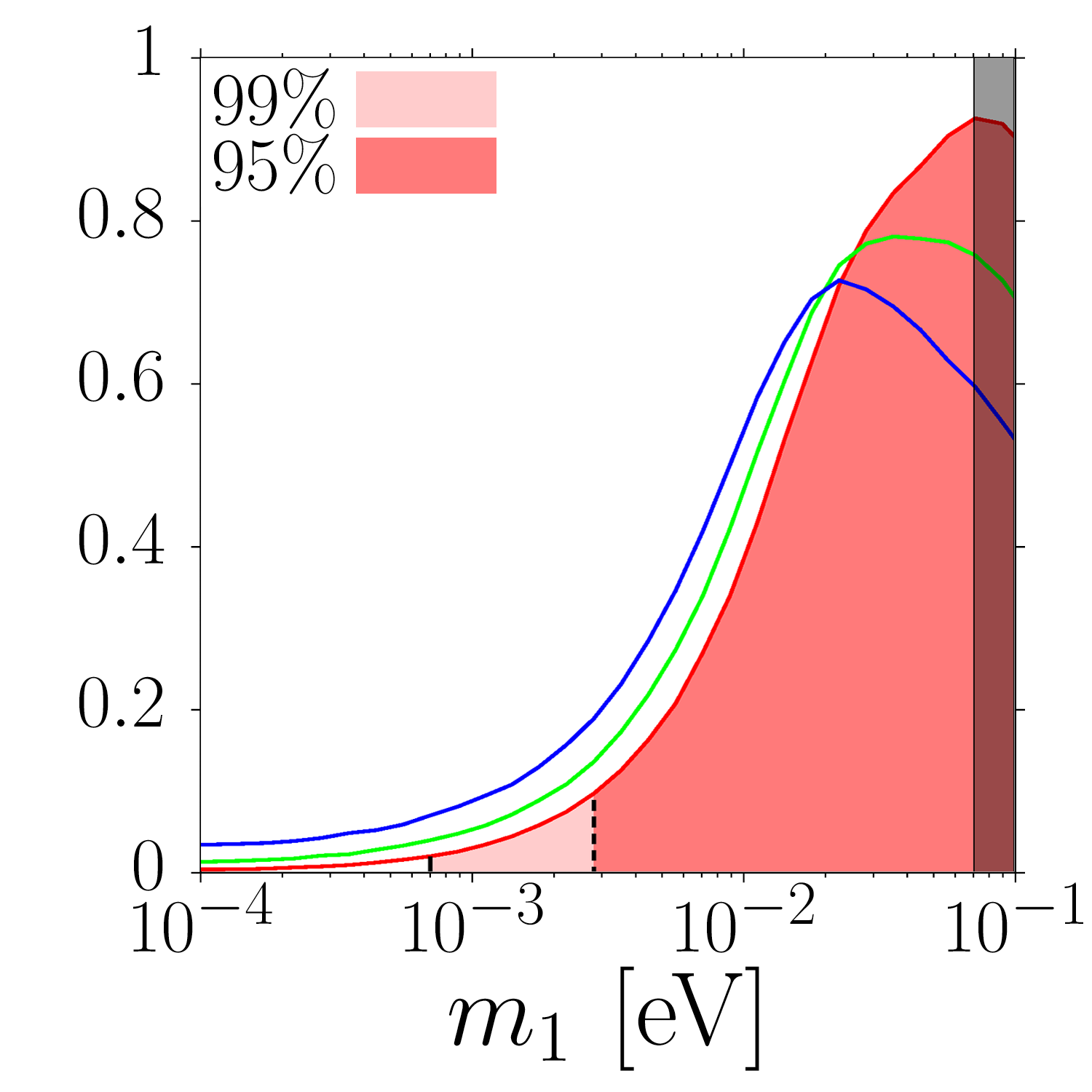,height=37mm,width=39mm}
\psfig{file=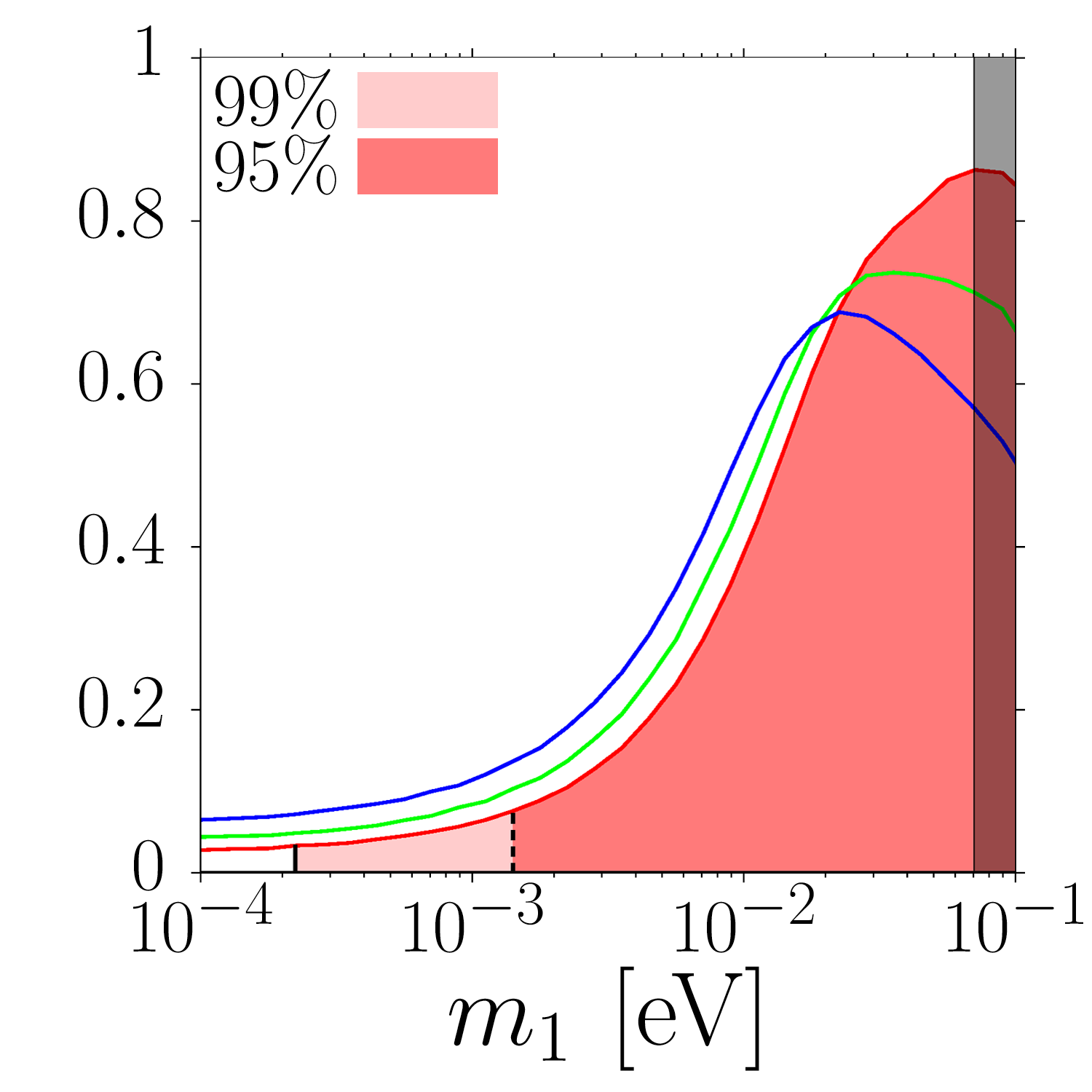,height=37mm,width=39mm}
\end{center}\vspace{-7mm}
\caption{IO case. Density of probabilities of $m_1$ from the scatter plots 
for $M_{\O}=1,2,5,10$ from left to right (same conventions as in Fig.~1).}
\label{densities}
\end{figure}
 One can see how this time there is no lower bound for $M_{\O}=1,2,5,10$ 
 and we could obtain points satisfying successful strong thermal leptogenesis 
 with arbitrarily small $m_1$. 
 
  However,   the fact that $K_{1\m}^{0,{\rm max}}$ is just slightly higher than $K_{\rm st}(N_{\D_\m})$
 (this time $K_{1\m}^{0,{\rm max}} \simeq 11\,M_{\O}$)
 still implies that one has to fine tune the parameters in the orthogonal  matrix in order to maximise
 $K_{1\m}$, and this still acts in a way that in the limit $m_1/m_{\rm atm} \ra 0$ the 
 density of points drops quickly.  
 For example one can see that for $M_{\O}=2$ one 
 still has that $99\%$ of the solutions are found for values $m_1 \gtrsim 3\,{\rm meV}$
 (the value quoted in the abstract).

 In Fig.~6 we also show again the results of the scatter plots in the planes $K_{1\a}-m_1$
 ($\a=e,\m,\t$). One can see how, while values of $K_{1e} \gg K_{\rm st} \sim 10-13$ can be found for arbitrarily 
 small values of $m_1$, the maximum value of $K_{1\mu}$  for small values of $m_1 \ll m_{\rm atm}$
 is just slightly greater than $K_{\rm st}$.  This confirms that $K_{1\mu}$ is the crucial quantity that constrains 
 $m_1$ in the case of IO, since the orthogonal matrix has to be strongly fine tuned in
 order to have $K_{1\m}\gtrsim K_{\rm st}$.   
 \begin{figure}
\begin{center}
\hspace{-5mm}
\psfig{file=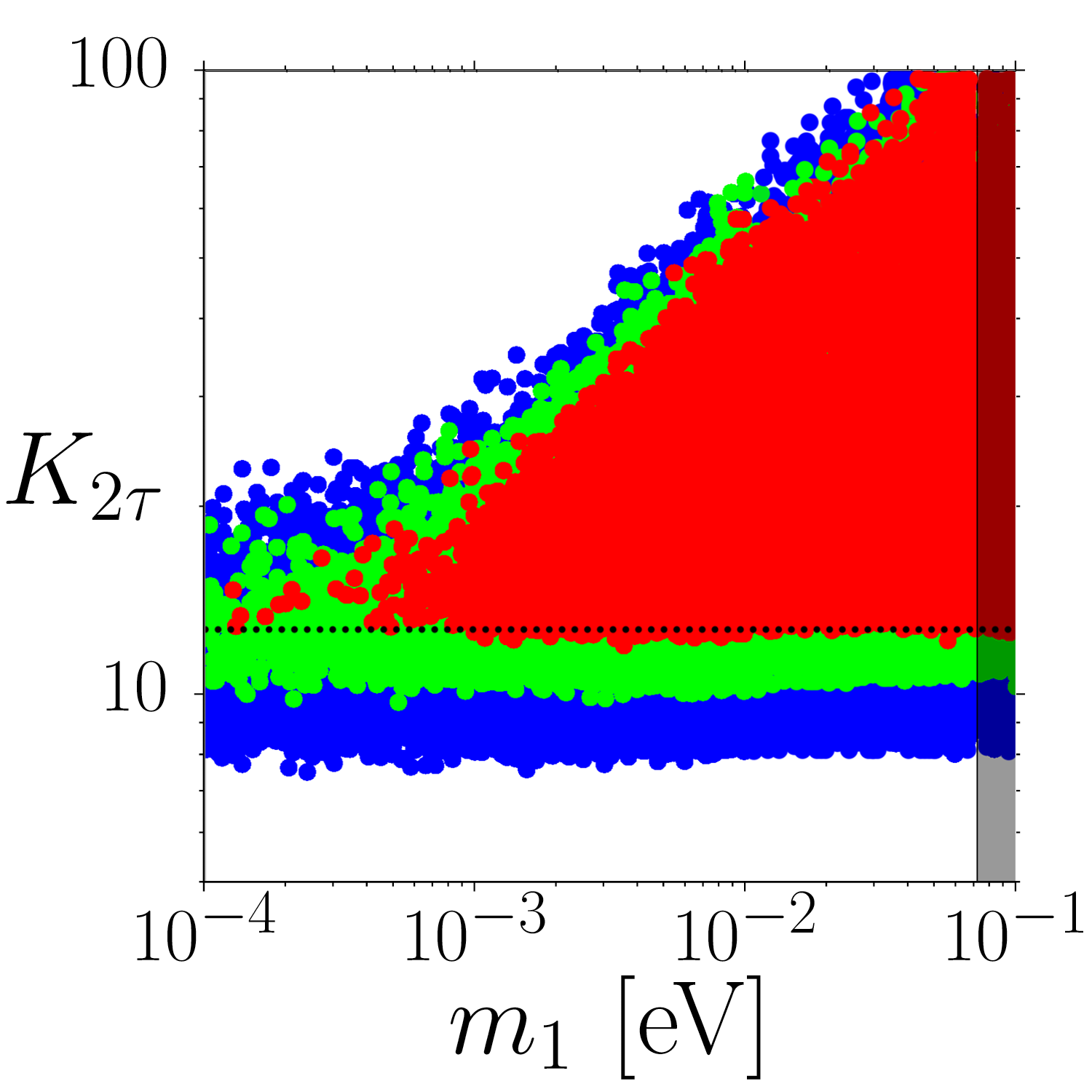,height=37mm,width=38mm}
\psfig{file=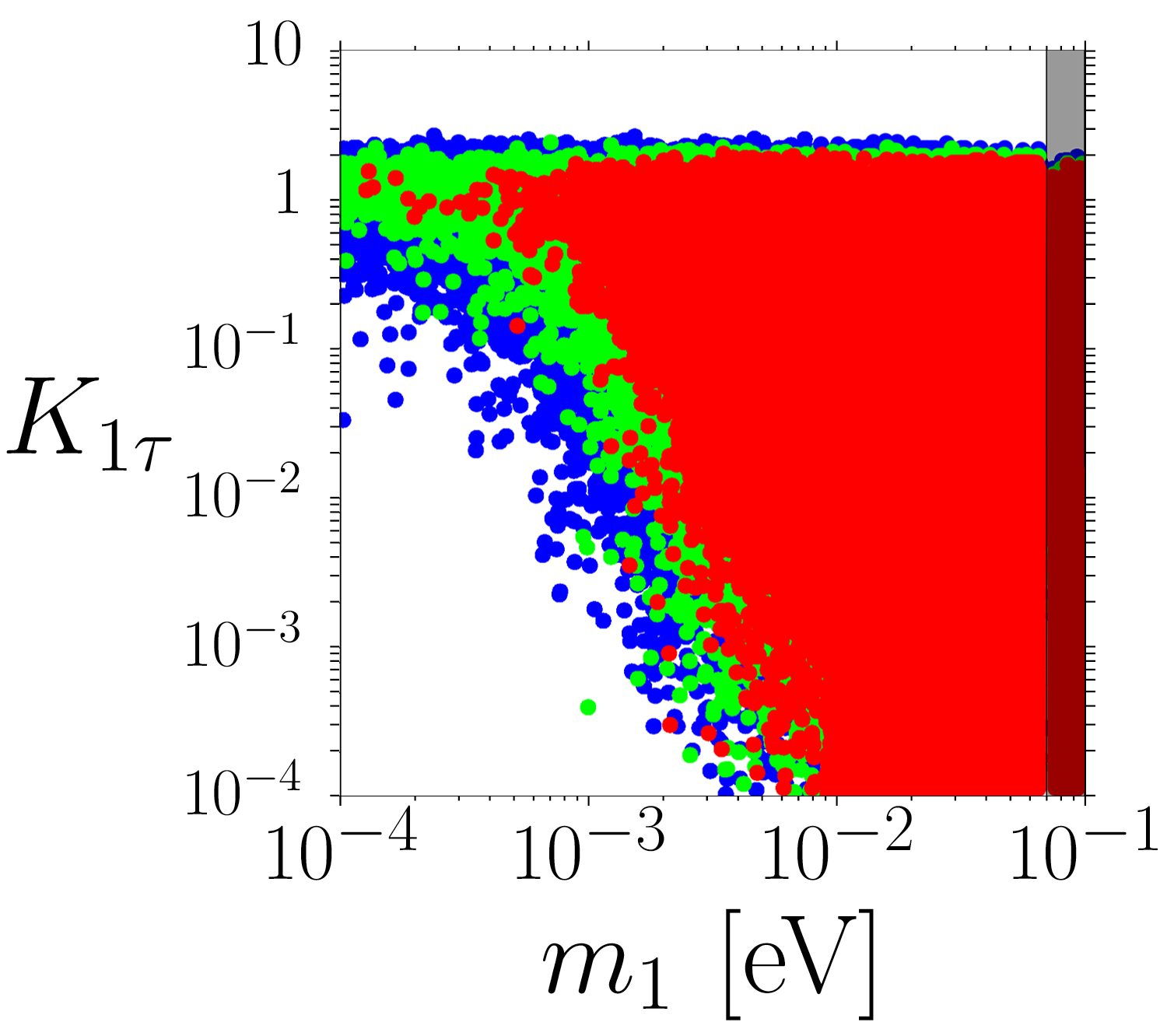,height=38mm,width=43mm}     
\psfig{file=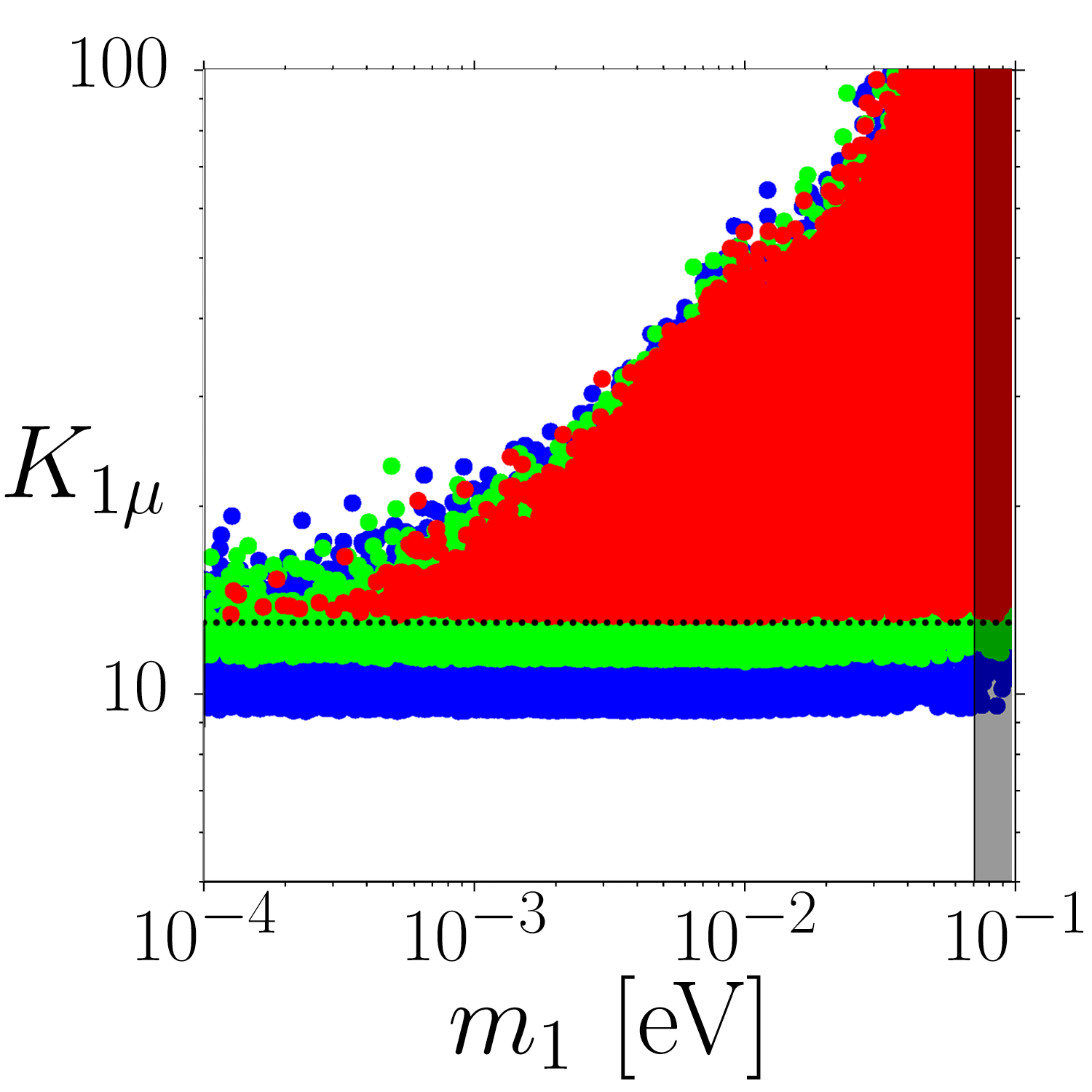,height=37mm,width=38mm}
\psfig{file=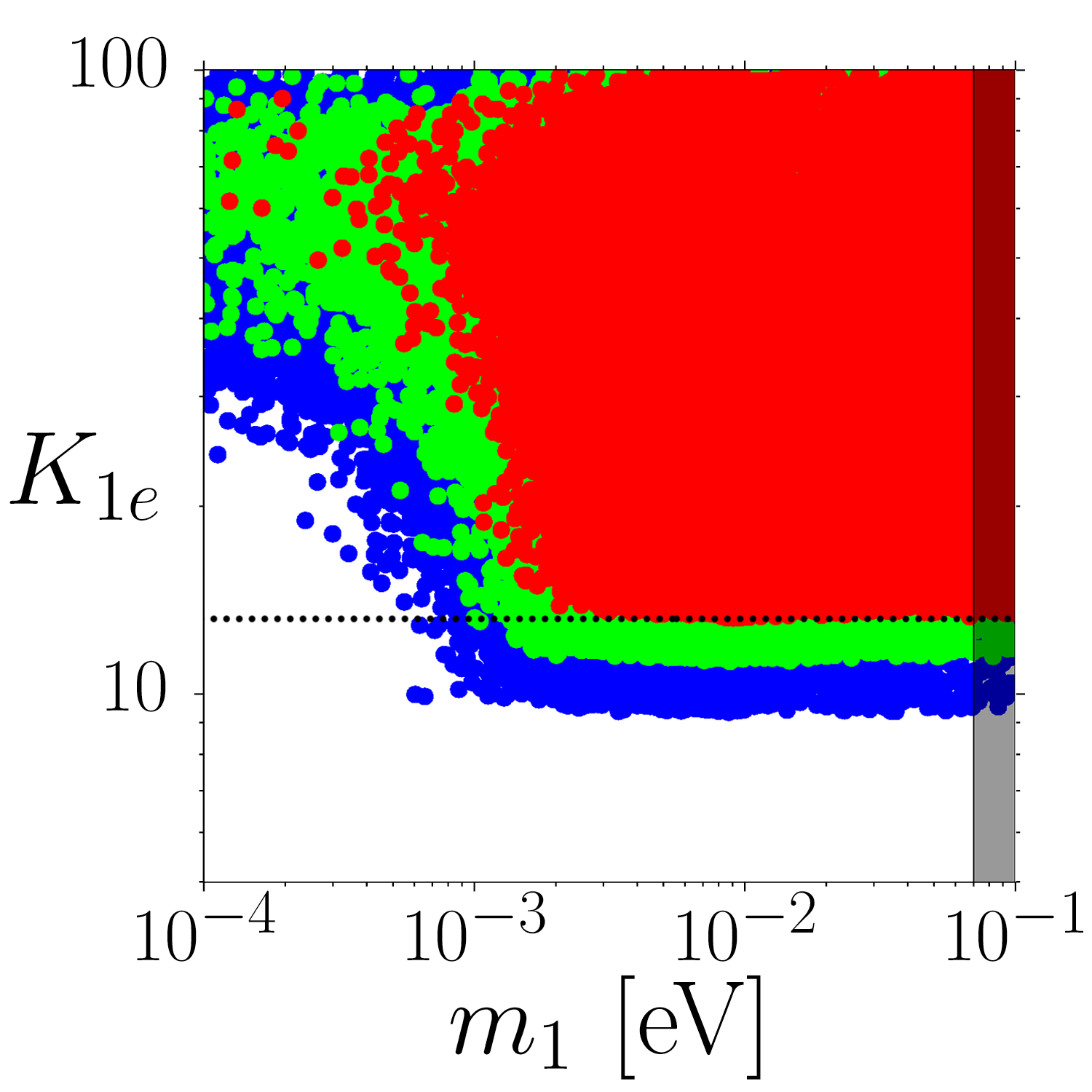,height=37mm,width=38mm}
\end{center}\vspace{-6mm}
\caption{IO case. Results of the scatter plots for $M_{\Omega}=2$ for the four relevant flavoured
decay parameters $K_{2\t}, K_{1\t}, K_{1\m}, K_{1 e}$ versus $m_1$ (same conventions as in Fig.~1).
The horizontal dotted line indicate $K_{\rm st}(\D N^{\rm p,i}_{\D\a}=0.03)$.}
\label{fdparametersIO}
\end{figure}

\subsection{Case $M_3 \lesssim 5\times 10^{11}\,{\rm GeV}$}

As pointed out in 3.2, for $M_3 \lesssim 5 \times 10^{11}\,$GeV, the condition 
$K_{2\tau} \gtrsim K_{\rm st}(N^{\rm p,i}_{\D_\t}) \gg 1$ gets relaxed into
$K_{2\tau}+K_{3\tau}\gg K_{\rm st}(N^{\rm p,i}_{\D_\t}) $. 
Potentially this condition can be much more easily satisfied and in particular
the value of $K_{2\tau}$ has not to be necessarily very large. In this way the condition of successful
leptogenesis becomes independent of the value of the initial pre-existing asymmetry and can be more easily 
satisfied. 

However, this point does not substantially change the results on the absolute neutrino mass scale
obtained for the case of large $M_3$.  The reason is that these, as we have seen, depend only on the
the $K_{1\a}$'s rather than on $K_{2\t}$ and in particular on the fact that for the NO (IO) case the value of
$K_{1e}^{0,{\rm max}}$ ($K_{1\m}^{0,{\rm max}}$) is very close to $K_{\rm st}$. 
In Fig.~7 we show again $K_{2\t}$ and the three $K_{1\a}$ for the NO case. 
 \begin{figure}
\begin{center}
\hspace{-5mm}
\psfig{file=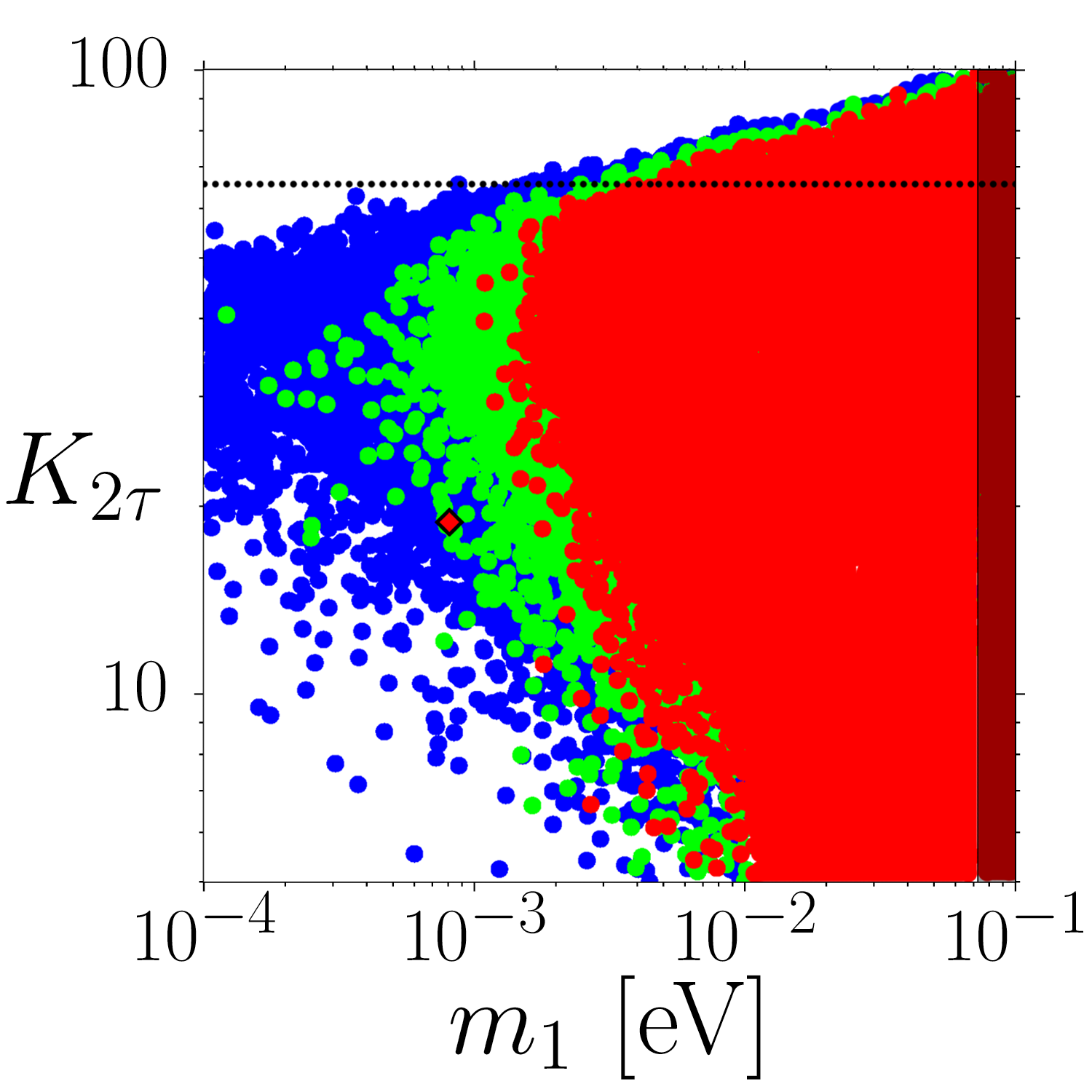,height=37mm,width=38mm}
\psfig{file=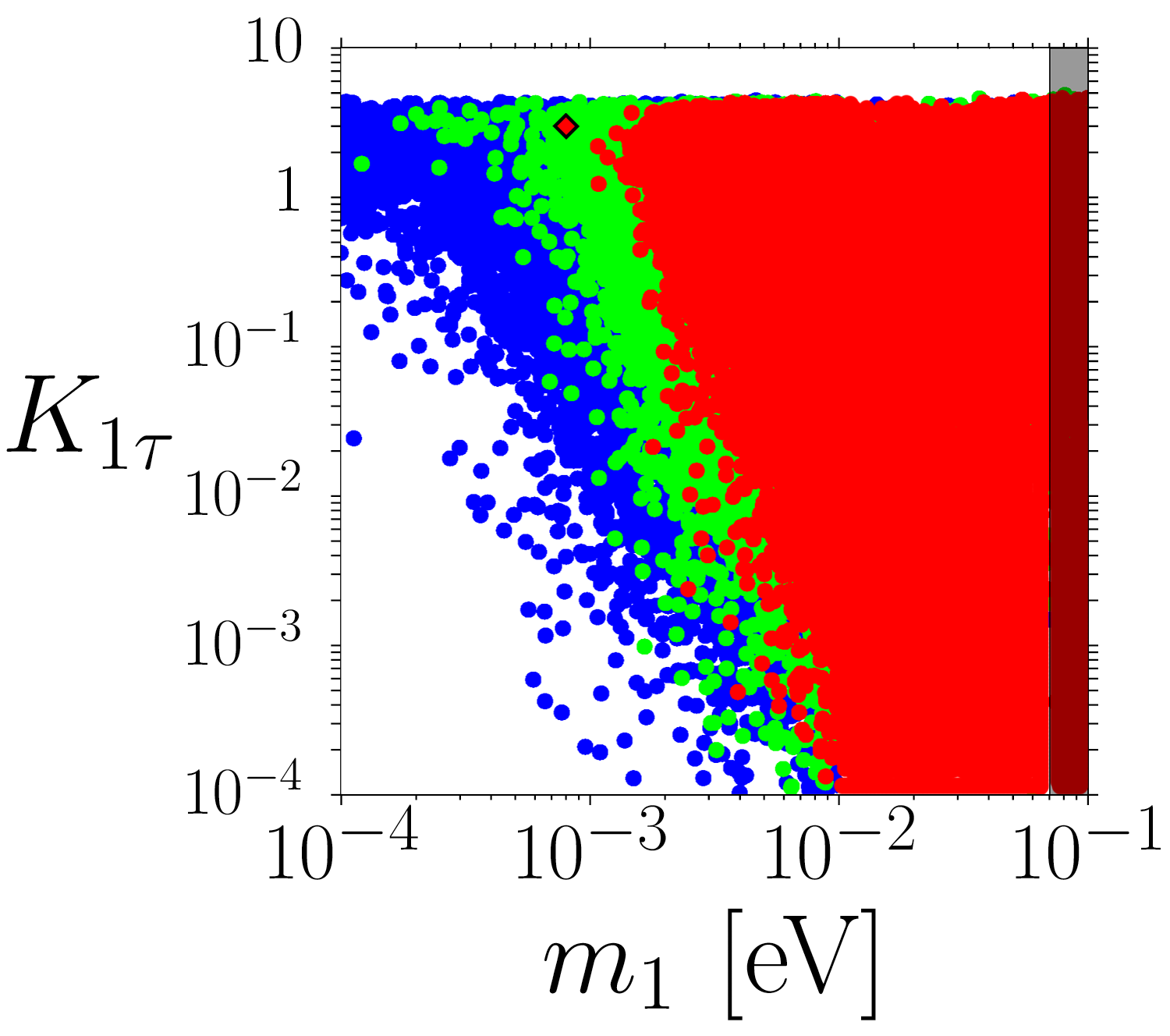,height=38mm,width=45mm}     
\psfig{file=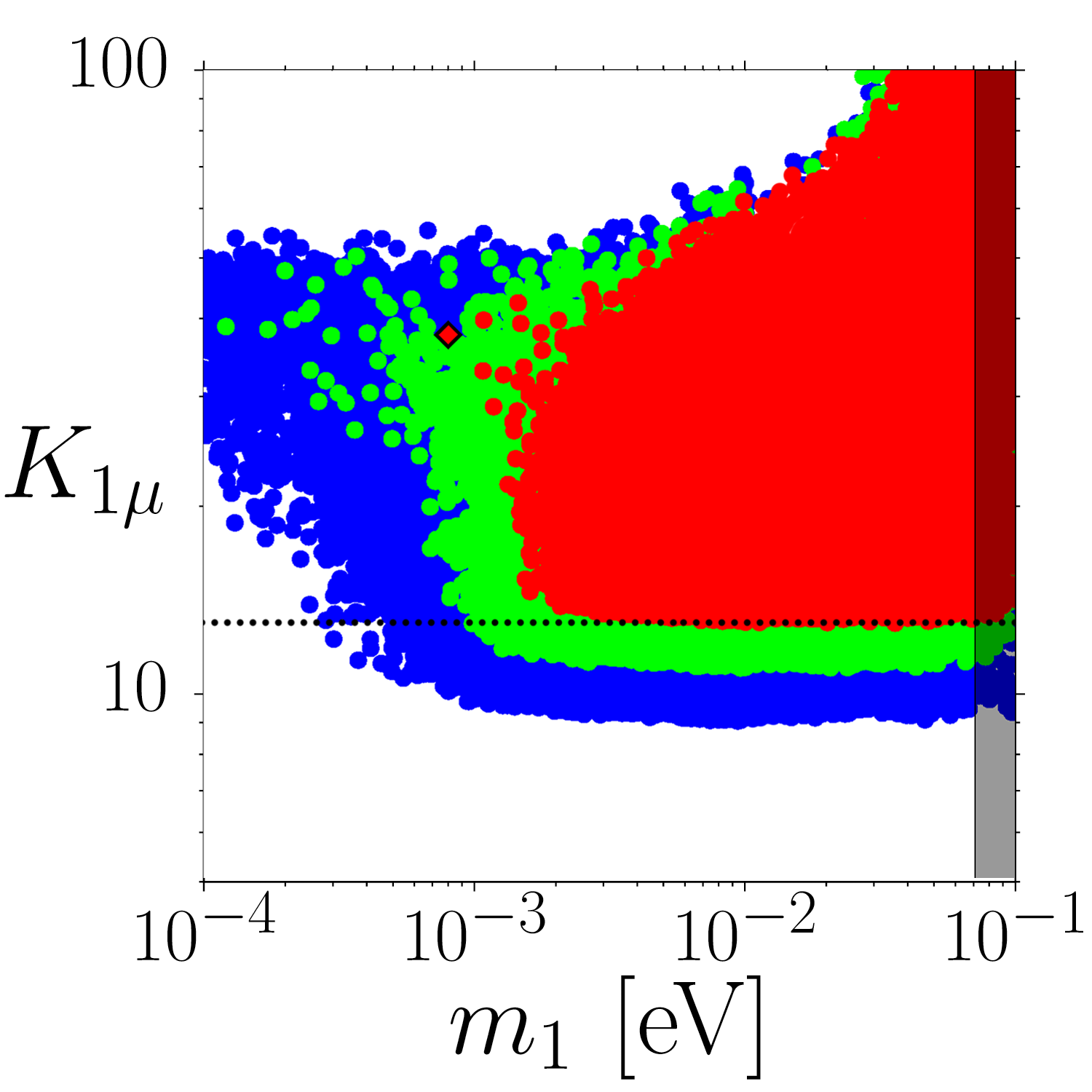,height=37mm,width=38mm}
\psfig{file=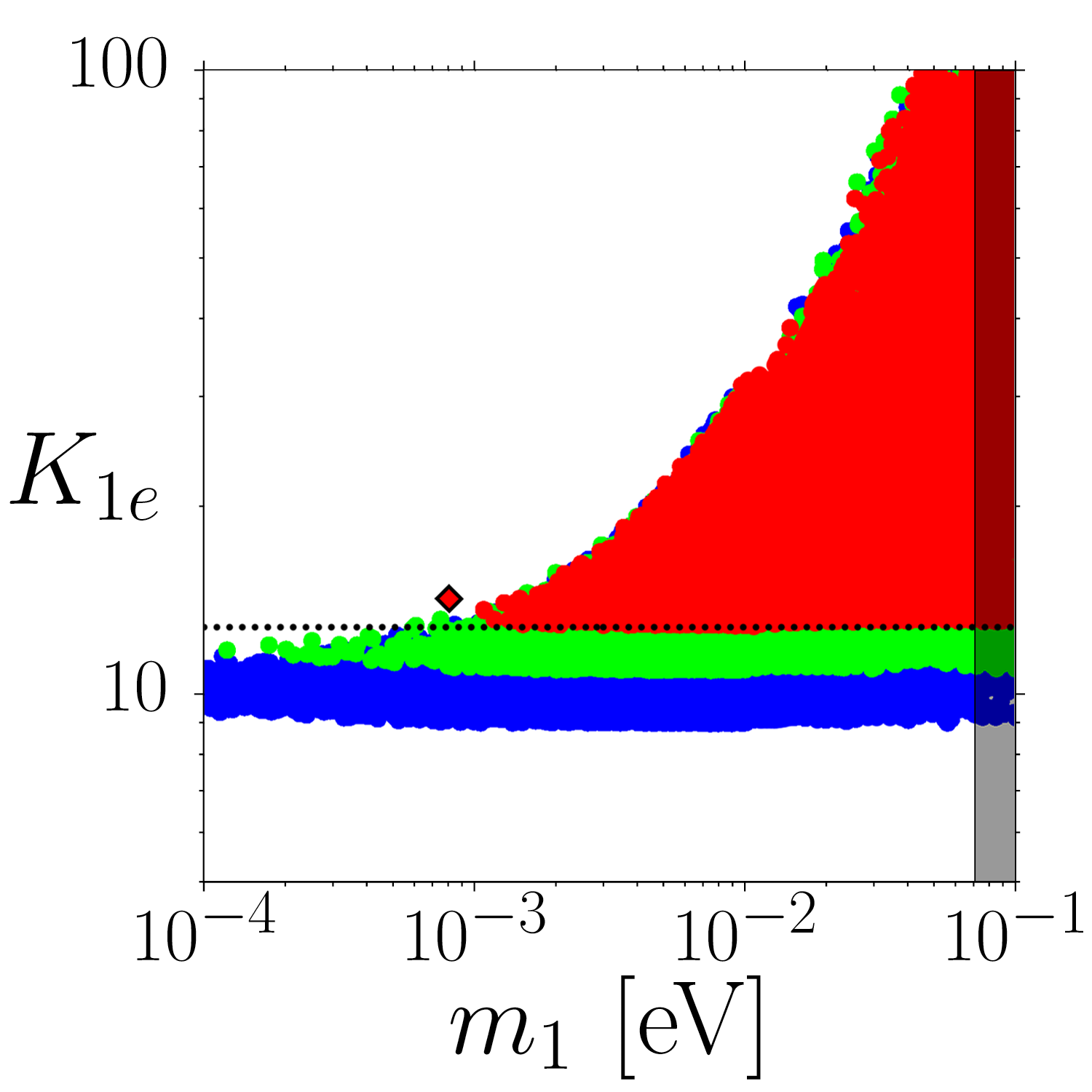,height=37mm,width=38mm}
\end{center}\vspace{-5mm}
\caption{NO case with $M_3 \lesssim 5\times 10^{11}\,{\rm GeV}$. 
Results of the scatter plots for $M_{\Omega}=2$ for the four relevant flavoured
decay parameters $K_{2\t}, K_{1\t}, K_{1\m}, K_{1e}$ versus $m_1$ (same conventions as in Fig.~1).
The horizontal dotted line indicate $K_{\rm st}(N^{\rm p,i}_{\D_{\a}}=0.03)$.}
\label{fdparametersNOlowM3}
\end{figure}
One can compare the results with those obtained for the case of large $M_3$ shown in Fig.~3 and notice how
except for $K_{2\t}$, that now can also be below $K_{\rm st}$, the scatter plot for $K_{1e}$, the crucial
quantity, is substantially the same. 

\subsection{A few comments on the results}

 Let us discuss a few points before concluding this section.

{\em The results depend on neutrino oscillation experimental data}. It should be noticed how the results
we obtained rely on the smallness of $K_{1e}^{0,{\rm max}}$ ($K_{1\m}^{0,{\rm max}}$) for NO (IO)
for $K_{1\t}\lesssim 1$ and this is enforced by the current measured value of the PMNS matrix entries
as we have seen, in particular $|U_{e3}|^2\ll 1$ for NO and 
$|U_{\mu 3}-U_{\tau 2}\,U_{\mu 3}/U_{\tau 3}|^2 \ll 1$ for IO. 
Therefore, the strong thermal leptogenesis condition realises an interesting interplay between
low energy neutrino data and leptogenesis predictions. 

{\em Theoretical uncertainties}. Our results have been derived using the analytical expressions
eqs.~(\ref{kappa}), (\ref{finalpas}) and (\ref{finalpaslowM3}). We have already noticed how these can be derived in the appropriate limit
from density matrix equations. Our results neglect momentum dependence in the wash-out but it has been noticed
that in the case of strong wash-out, as imposed by strong thermal leptogenesis, this approximation
underestimates the wash-out \cite{pastorvives} though it has been recently claimed that this is actually an effect
that arises not from momentum dependence but from a proper account of 
quantum statistics in the wash-out rates that increase them by $20\%$ \cite{bodeker}.
This would tend to slightly relax our lower bound. On the other hand, taking into account
Higgs and quarks asymmetries, would act in the opposite direction \cite{fuller}. Another consequence of accounting for 
these asymmetries is flavour coupling. This would tend to open new ways to the pre-existing asymmetry to escape the lightest RH neutrino
wash-out \cite{fuller}. Account of flavour coupling would then act into the direction of tightening the lower bound
and this is likely the strongest effect. These effects will be taken into account in a forthcoming publication. 

{\em Case $\D p_{{\rm p} \a}=0$}. How do the results change if the pre-existing 
asymmetry is assumed to have the same flavour composition for leptons and anti-leptons, 
so that $\D p_{{\rm p} \a}=0$ in the eqs.~(\ref{finalpas}), (\ref{finalpaslowM3})? 
In this case there is no lower bound
for any value of $M_{\O}$, simply because now the strong thermal condition 
is also satisfied if $(1-p^0_{e\tau^{\bot}_2}) \lesssim 10^{-7}$,
independently of the value of $K_{1e}$ depending on $m_1$. 
However, it is clear that this possibility is realised for very special models where 
basically the $N_2$'s have to decay  into leptons without a muon component, i.e. $K_{2\mu}=0$, 
a very special case though not excluded by experimental data.  
Indeed in the scatter plots we find a few of such points independently of $m_1$.
However,  even though they evade the lower bound on $m_1$,  
they basically do not modify the $m_1$ distributions. Therefore, this caveat
corresponds to a very special and definite situation that does not change the general results.  

{\em $SO(10)$-inspired models}.  Our results are in perfect agreement with the 
results found in \cite{strongSO10lep} where, in addition to the strong thermal condition,
$SO(10)$-inspired conditions are also imposed on the Dirac neutrino mass matrix. In this
case NO case is a necessary condition. Moreover one finds $M_{\O}\lesssim 0.8$ and our lower bound gives $m_1 \gtrsim 10\,{\rm meV}$
that is indeed respected since the range $m_1 = (15\mbox{--}25)\,{\rm meV}$ is found, showing that the $SO(10)$-inspired
conditions further restrict $m_1$ basically pinning down a very narrow range for $m_1$.

{\em Form-dominance models \cite{king}}.   In these models each light neutrino mass
is inversely proportional to one different RH neutrino mass. They correspond to 
an orthogonal matrix equal to one of the six permutation matrices \cite{geometry}. In this situation the needed 
cancellation in the eq.~(\ref{condition}), in order to have $K_{1\tau}\lesssim 1$ 
and at the same time large $K_{1e}$ and $K_{1\m}$, is impossible.
The only way to have a small $K_{1\t}$ in these models is to have small $m_1$ values with  $|\O_{11}|\simeq 1$
(while necessarily $\O_{21}, \O_{31} \simeq 0$) but in this case then, as we have seen, one cannot simultaneously satisfy the conditions $K_{1e}, K_{1\m} \gg 1$. Therefore form-dominance models 
cannot realise  strong thermal  leptogenesis.  These two examples show how our analytical procedure can 
be applied to specific models with definite and in general much stronger constraints on $m_1$.

\subsection{Prospects from future experiments}

\subsubsection{The importance of solving the ambiguity on neutrino mass ordering}
 
As we have seen for NO successful strong thermal leptogenesis 
favours $m_1\gtrsim 10\,{\rm meV}$ for $M_{\O}\lesssim 2$ and in our scatter
plots we found less than $1\%$ of points at lower values.
There is even a strict lower bound $m_1 \gtrsim 1\,{\rm meV}$
valid for any choice of the orthogonal matrix.  For IO the constrains are looser.
There is not such a  strict lower bound and only for $m_1 \lesssim 3\,{\rm meV}$
we found  a number of points less than $1\%$.  It is then very important that in the next years 
 neutrino oscillations experiments will be able to solve the ambiguity between NO and IO
 neutrino masses. If NO will prove to be correct, then strong thermal leptogenesis 
 can be more easily tested since it strongly favours $m_1 \gtrsim 10\,{\rm meV}$, 
 values sufficiently large to produce measurable deviations 
 from the full hierarchical case (i.e. semi-hierarchical neutrinos)
 in cosmological observations. 

\subsubsection{Cosmological observations}

Cosmological observations are sensitive to neutrino masses and are
able to place an upper bound, typically quoted on $\sum_i \, m_i$ (though future observations might 
become  sensitive to the full neutrino spectrum). Future observations could potentially reach a precision of 
$\d (\sum_i m_i) \simeq 10\,{\rm meV}$ \cite{hannestad}. 
In the case of NO, assuming that they would be able to measure the hierarchical lower limit 
finding $\sum_i m_i = (60 \pm 10)\,{\rm meV}$, they would be able to place a 
$2\s$ upper bound $m_1 \lesssim 10\,{\rm meV}$.  From our results this means that future cosmological 
observations will be potentially able to severely constraint strong thermal leptogenesis with 
hierarchical RH neutrinos. 
On the other hand a measurement $\sum_i m_i \gtrsim (95 \pm 10) \,{\rm meV}$ would correspond to 
$m_1 \gtrsim (20 \pm 5)\,{\rm meV}$, allowing to place a $2\,\s$ lower bound $m_1 \gtrsim 10\,{\rm meV}$,
and this would be in agreement with the expectations from strong thermal leptogenesis.
In the case of IO, a measurement $\sum_i m_i = (100 \pm 10)\,{\rm meV}$, 
in agreement with the hierarchical limit for IO, 
 would correspond to a $2\s$ upper bound $m_1 \lesssim 15\,{\rm meV}$,
representing a much looser constraint on strong thermal leptogenesis that in the NO case. 
Moreover expected values $m_1 \gtrsim 3\,{\rm meV}$ would correspond to 
 measurements $\sum_i m_i \gtrsim (100\pm 10)\,{\rm meV}$, in general not distinguishable
 from the inverted hierarchical limit, i.e. not testable.    
This shows how NO would be a much more favourable option than IO for a significant test (negative or positive)
of strong thermal leptogenesis, since it more strongly favours detectable deviations from 
the hierarchical limit ($m_1 \ra 0$).  It should be noticed that NO ordered neutrino masses with $m_1 \simeq 20\,{\rm meV}$
would also yield $\sum_i m_i \simeq 100\,{\rm meV}$ as for IO hierarchical neutrino masses ($m_1 \ll m_{\rm sol}$)
and this is another reason why it is important that neutrino oscillation experiments will be able to solve the NO-IO ambiguity
independently of absolute neutrino mass experiments. 

\subsubsection{Neutrinoless double beta decay experiments}

In the central panel of Fig.~1 we have also plotted the values of the neutrinoless double 
beta decay effective neutrino mass $m_{ee}$ versus $m_1$ from the scatter plot (both for NO and IO). 
We have also shown the results without imposing
strong thermal leptogenesis (yellow points).  It can be seen how for NO, since the effective neutrino mass
can be well below $m_1$ thanks to phase cancellations \cite{rodejohann}, this can be as small
as $\sim 1\,{\rm meV}$ even for $m_1 \gtrsim 10\,{\rm meV}$ (as indicated by the horizontal and vertical 
solid lines respectively). 
This implies that strong thermal leptogenesis is not able to produce effective constraints on $m_{ee}$.
Vice-versa, however, a future measurement of $m_{ee} \gtrsim 10\,{\rm meV}$ would imply necessarily
$m_1 \gtrsim 10\,{\rm meV}$ providing an interesting strong support 
to the strong thermal leptogenesis expectations .  For IO, again, the strong thermal prediction hardly
produces detectable deviations from the inverted hierarchical limit. 

\subsubsection{Tritium beta decay experiments}

In the case of absence of signal,  the KATRIN experiment will be able to place 
an upper bound onto the effective electron neutrino mass $m_{\nu_e} \lesssim 250 \,{\rm meV}$
translating into a similar upper bound on $m_1$. Therefore, it will not be able to place severe constraints
on strong thermal leptogenesis. In the PROJECT 8 experimental proposal \cite{project8}, the energy of electrons 
emitted in Tritium beta decay is determined from the frequency of cyclotron radiation and
the upper bound could be improved to $m_{\nu_e} \lesssim 50\,{\rm meV}$. This would translate again into
a similar upper bound on $m_1$, providing a more stringent constraint but still not able 
to severely corner strong thermal leptogenesis.

\section{Conclusions}

Thanks to the current measured values of the neutrino mixing angles, and in particular of $\theta_{13}$,
the assumption of strong thermal leptogenesis can be tested quite strongly by future
cosmological observations, especially in the NO case. If these will be able to place a stringent 
upper bound on the lightest neutrino mass scale $m_1 \lesssim 10\,{\rm meV}$, then 
they will strongly corner the idea of strong thermal leptogenesis. This will survive only admitting 
quite a strong fine tuning in the seesaw formula and/or in the flavoured decay parameters. 
The result would be much stronger for the NO case than the for the IO case. Therefore, it is
important that future neutrino oscillation experiments will be able to solve the NO-IO ambiguity. 
On the other hand a positive measurement  $m_1 \gtrsim 10\,{\rm meV}$ 
could be certainly considered as an important experimental information supporting
strong thermal leptogenesis.  
It is  fascinating that, thanks to the forthcoming advance in the
determination of neutrino parameters, we will have soon the opportunity to test 
important theoretical ideas  in relation to a fundamental cosmological puzzle 
such as the observed matter-antimatter asymmetry of the Universe. 

\vspace{-1mm}
\subsection*{Acknowledgments}

We thank Luca Marzola  for useful comments and discussions. We acknowledge computer resources
from the IRIDIS High Performance Computer facility (University of Southampton).
PDB and SK acknowledge financial support  from the NExT/SEPnet Institute.
PDB   acknowledges financial support also from the STFC Rolling Grant ST/G000557/1 and from the  
EU FP7  ITN INVISIBLES  (Marie Curie Actions, PITN- GA-2011- 289442).
MRF acknowledges financial support from the STAG Institute.


\begin{thebibliography}{99}

\bibitem{planck}
  P.~A.~R.~Ade {\it et al.}  [Planck Collaboration],
  arXiv:1303.5076 [astro-ph.CO].

\bibitem{fy}
M.~Fukugita, T.~Yanagida, \pl{174}{1986}{45}.

 \bibitem{seesaw}
P.~Minkowski, Phys.\ Lett.\ B {\bf 67} (1977) 421;
T.~Yanagida, in {\it{Workshop on Unified Theories}}, KEK report
79-18 (1979) p.~95;
M.~Gell-Mann, P.~Ramond, R.~Slansky, in {\it{Supergravity}} (North Holland,
Amsterdam, 1979) eds. P.~van Nieuwenhuizen, D.~Freedman, p.~315;
S.L. Glashow, in {\it 1979 Cargese Summer Institute on Quarks and Leptons}
(Plenum Press, New York, 1980) 
p.~687;
R.~Barbieri, D.~V.~Nanopoulos, G.~Morchio and F.~Strocchi,
Phys.\ Lett.\ B {\bf 90} (1980) 91;
R.~N.~Mohapatra and G.~Senjanovic,
Phys.\ Rev.\ Lett.\  {\bf 44} (1980) 912.

\bibitem{nardini}
T.~Cohen, D.~E.~Morrissey and A.~Pierce,
  Phys.\ Rev.\ D {\bf 86} (2012) 013009;
D.~Curtin, P.~Jaiswal and P.~Meade,
  JHEP {\bf 1208} (2012) 005;  
M.~Carena, G.~Nardini, M.~Quiros and C.~E.~M.~Wagner,
  JHEP {\bf 1302} (2013) 001.

\bibitem{review}
For a recent review  see
S.~Blanchet and P.~Di Bari, 
  New J.\ Phys.\  {\bf 14} (2012) 125012.

\bibitem{grav}
 R.~Kallosh, A.~D.~Linde, D.~A.~Linde and L.~Susskind,
  Phys.\ Rev.\  D {\bf 52} (1995) 912;
 H.~Davoudiasl, R.~Kitano, G.~D.~Kribs, H.~Murayama and P.~J.~Steinhardt,
  Phys.\ Rev.\ Lett.\  {\bf 93} (2004) 201301.


\bibitem{GUT}
 M.~Yoshimura,
  Phys.\ Rev.\ Lett.\  {\bf 41} (1978) 281
  [Erratum-ibid.\  {\bf 42} (1979) 746];
 S.~Dimopoulos and L.~Susskind,
  Phys.\ Rev.\  D {\bf 18} (1978) 4500;
  D.~Toussaint, S.~B.~Treiman, F.~Wilczek and A.~Zee,
  Phys.\ Rev.\  D {\bf 19} (1979) 1036;
E.~W.~Kolb and S.~Wolfram,
  Nucl.\ Phys.\  B {\bf 172} (1980) 224
  [Erratum-ibid.\  B {\bf 195} (1982) 542].
 E.~W.~Kolb, A.~D.~Linde and A.~Riotto,
  Phys.\ Rev.\ Lett.\  {\bf 77} (1996) 4290.

\bibitem{AD}
I.~Affleck and M.~Dine,
  Nucl.\ Phys.\  B {\bf 249} (1985) 361.

\bibitem{valle}
D.~V.~Forero, M.~Tortola and J.~W.~F.~Valle,
  Phys.\ Rev.\ D {\bf 86} (2012) 073012.

\bibitem{newfogli}
F.~Capozzi, G.~L.~Fogli, E.~Lisi, A.~Marrone, D.~Montanino and A.~Palazzo,
  arXiv:1312.2878 [hep-ph].
  
 \bibitem{gonzalez}
M.~C.~Gonzalez-Garcia, {\rm et al}
  JHEP {\bf 1212} (2012) 123.


\bibitem{window}
W.~Buchmuller, P.~Di Bari and M.~Plumacher,
  Nucl.\ Phys.\ B {\bf 665} (2003) 445.

\bibitem{problem}
  E.~Bertuzzo, P.~Di Bari, L.~Marzola,
  Nucl.\ Phys.\  {\bf B849 } (2011)  521-548.



\bibitem{strumianardinir}
R.~Barbieri, P.~Creminelli, A.~Strumia and N.~Tetradis,
  Nucl.\ Phys.\ B {\bf 575} (2000) 61;
 G.~Engelhard, Y.~Grossman, E.~Nardi and Y.~Nir,
  Phys.\ Rev.\ Lett.\  {\bf 99} (2007) 081802.


 \bibitem{geometry}
P.~Di Bari,
 Nucl.\ Phys.\  {\bf B727 } (2005)  318-354.
 
\bibitem{densitymatrix} 
A.~Abada, S.~Davidson, F.~-X.~Josse-Michaux, M.~Losada and A.~Riotto,
  JCAP {\bf 0604} (2006) 004;
S.~Blanchet, P.~Di Bari and G.~G.~Raffelt,
  JCAP {\bf 0703} (2007) 012;
 A.~De Simone and A.~Riotto,
  JCAP {\bf 0702} (2007) 005;
M.~Beneke, B.~Garbrecht, C.~Fidler, M.~Herranen and P.~Schwaller,
  Nucl.\ Phys.\ B {\bf 843} (2011) 177;
B.~Garbrecht, F.~Glowna and P.~Schwaller,
  Nucl.\ Phys.\ B {\bf 877} (2013) 1. 
 
\bibitem{N2dominated}
 O.~Vives,
  Phys.\ Rev.\  {\bf D73 } (2006)  073006;
  S.~Blanchet and P.~Di Bari,
  Nucl.\ Phys.\ B {\bf 807} (2009) 155;
P.~Di Bari and A.~Riotto,
  Phys.\ Lett.\ B {\bf 671} (2009) 462.
  
\bibitem{fuller}  
 S.~Antusch, P.~Di Bari, D.~A.~Jones and S.~F.~King,
  Nucl.\ Phys.\ B {\bf 856} (2012) 180.
  
  \bibitem{densitymatrix2}
  S.~Blanchet, P.~Di Bari, D.~A.~Jones and L.~Marzola,
  JCAP {\bf 1301} (2013) 041.
  
  \bibitem{pedestrians} 
W.~Buchmuller, P.~Di Bari and M.~Plumacher,
  Annals Phys.\  {\bf 315} (2005) 305.

\bibitem{flavorlep}
  S.~Blanchet and P.~Di Bari,
  JCAP {\bf 0703} (2007) 018.

\bibitem{crv}  
L.~Covi, E.~Roulet and F.~Vissani,
  Phys.\ Lett.\ B {\bf 384} (1996) 169.
 
\bibitem{casas}
J.~A.~Casas and A.~Ibarra,
  Nucl.\ Phys.\ B {\bf 618} (2001) 171.

\bibitem{strongSO10lep}
P.~Di Bari and L.~Marzola,
  Nucl.\ Phys.\ B {\bf 877} (2013) 719.

\bibitem{2RHN}
S.~Antusch, P.~Di Bari, D.~A.~Jones and S.~F.~King,
  Phys.\ Rev.\ D {\bf 86} (2012) 023516.

\bibitem{hambye}
  T.~Hambye, Y.~Lin, A.~Notari, M.~Papucci and A.~Strumia,
  Nucl.\ Phys.\ B {\bf 695} (2004) 169;
 S.~Blanchet and P.~Di Bari,
  Nucl.\ Phys.\ B {\bf 807} (2009) 155.  

\bibitem{pastorvives}
J.~Garayoa, S.~Pastor, T.~Pinto, N.~Rius and O.~Vives,
  JCAP {\bf 0909} (2009) 035.
  
\bibitem{bodeker}
  D.~Bodeker and M.~Wormann,
  JCAP02(2014)016.
  
  

\bibitem{king}
M.~C.~Chen and S.~F.~King,
  JHEP {\bf 0906} (2009) 072.

\bibitem{hannestad}
J.~Hamann, S.~Hannestad and Y.~Y.~Y.~Wong,
  JCAP {\bf 1211} (2012) 052.

\bibitem{rodejohann}
W.~Rodejohann,
 Int.\ J.\ Mod.\ Phys.\  E {\bf 20} (2011) 1833.

\bibitem{project8}
P.~J.~Doe {\it et al.}  [Project 8 Collaboration],
  arXiv:1309.7093 [nucl-ex].


\end{thebibliography}
\end{document}